\documentclass[usenatbib,fleqn]{aa}  

\usepackage{graphicx}
\usepackage{txfonts}
\usepackage{amsmath}	
\usepackage{amssymb}	
\usepackage{natbib}
\usepackage{caption}
\usepackage{subcaption}
\usepackage[Symbol]{upgreek}
\usepackage{epstopdf}
\usepackage{float}
\usepackage{color}
\usepackage[colorlinks,citecolor=blue,linkcolor=blue,anchorcolor=blue]{hyperref}

\DeclareMathAlphabet{\pazocal}{OMS}{zplm}{m}{n}

\newcommand{\rg}{GM/c^2}
\newcommand{\tunit}{GM/c^3}

\newcommand{\trat}{T_{\rm p}/T_{\rm e}}

\graphicspath{{Figures/}}

\begin{document} 

\titlerunning{RAPTOR I}
\title{RAPTOR I: Time-dependent radiative transfer in arbitrary spacetimes\thanks{The public version of {\tt RAPTOR} is available at the following URL: https://github.com/tbronzwaer/raptor}}

\authorrunning{T. Bronzwaer et al.}
\author{T. Bronzwaer
\inst{1},
J. Davelaar\inst{1},\\
Z. Younsi\inst{2},
M. Mo{\'s}cibrodzka\inst{1},
H. Falcke\inst{1},
M. Kramer\inst{3},
\and L. Rezzolla\inst{2,4}
}

\institute{Department of Astrophysics/IMAPP, Radboud University Nijmegen
                  P.O. Box 9010, 6500 GL Nijmegen, The Netherlands
\and
Institut f\"ur Theoretische Physik, Max-von-Laue-Stra{\ss}e 1, D-60438
Frankfurt am Main, Germany
\and 
Max-Planck-Institut f\"ur Radioastronomie, Auf dem H\"ugel 69, Bonn
53121, Germany
\and
Frankfurt Institute for Advanced Studies,
  Ruth-Moufang-Stra{\ss}e 1, 60438 Frankfurt, Germany
}


   
\abstract
%
{Observational efforts to image the immediate environment of a black hole
  at the scale of the event horizon benefit from the development of
  efficient imaging codes that are capable of producing synthetic data,
  which may be compared with observational data. }
{We aim to present {\tt RAPTOR}, a new public code that produces accurate
  images, animations, and spectra of relativistic plasmas in strong
  gravity by numerically integrating the equations of motion of light
  rays and performing time-dependent radiative transfer calculations
  along the rays. The code is compatible with any analytical or numerical
  spacetime. It is hardware-agnostic and may be compiled and run both on
  GPUs and CPUs.}
{We describe the algorithms used in {\tt RAPTOR} and test the code's performance. We have performed a detailed comparison of {\tt
    RAPTOR} output with that of other radiative-transfer codes and demonstrate convergence of the results. We then applied {\tt RAPTOR} to
  study accretion models of supermassive black holes, performing
  time-dependent radiative transfer through general relativistic
  magneto-hydrodynamical (GRMHD) simulations and investigating the
  expected observational differences between the so-called fast-light and
  slow-light paradigms.}
{Using {\tt RAPTOR} to produce synthetic images and light curves of
  a GRMHD model of an accreting black hole, we find that the relative
  difference between fast-light and slow-light light curves is less than
  5\%. Using two distinct radiative-transfer codes to process the
  same data, we find integrated flux densities with a relative difference
  less than 0.01\%.}
{For two-dimensional GRMHD models, such as those examined in this
  paper, the fast-light approximation suffices as long as errors of a few percent are acceptable. The convergence of the results
  of two different codes demonstrates that they are, at a minimum,
  consistent.}

\keywords{radiative transfer -- black hole physics -- accretion,
  accretion disks}

\maketitle


\section{Introduction}

Testing Einstein's general theory of relativity (GR) in the strong-field
limit remains a difficult challenge. Recently, gravitational waves have
been used to test GR's predictions concerning merging black holes (see,
e.g., \citet{GWpaper2016}), whose observations are consistent with GR and
the existence of black holes \citep{Chirenti2016}. Building on
\citet{bardeen1972b}, \citet{falcke2000} suggested that a black hole's
`shadow' may be used to probe the accuracy of GR in the strong-field
limit; observational efforts using millimeter-wavelength very long
baseline interferometry (mm-VLBI) techniques are now underway
(\citealt{falcke2000}; \citealt{doeleman2008}; \citealt{doeleman2012};
\citealt{fish2016}; \citealt{johnson2015}; \citealt{goddi2016}).

Prime motivating targets for this research are the putative accreting
supermassive black holes (SMBHs) at the center of the Milky Way
(Sagittarius A*, hereafter Sgr A*) (\citealt{falckemarkoff};
\citealt{genzel2010}) and M87 \citep{doeleman2012}. The first attempts to
predict the electromagnetic appearance of black holes were made in the
1970's (\citealt{bardeen1972b}; \citealt{luminet1979};
\citealt{viergutz1993}), however, due to the limited observational
capacity of that era, it was considered a somewhat academic
exercise. With the advent of mm-VLBI techniques, and the potential to
image the black-hole shadows in Sgr A*, efficient and flexible GR
ray-tracing codes are becoming a key component of image-based tests of GR
in the strong-field limit. It is no longer sufficient to calculate the
appearance of thin disks or background stars, but rather, one has to
consider a black hole surrounded by a heterogeneous, partially optically
thick plasma. In order to accurately reproduce the appearance and
spectrum of such a source to a distant observer, the effects of a strong
gravitational field (gravitational lensing, redshift, and relativistic
boosting) must be taken into account. For these reasons, creating
synthetic observational data from physical models of the SMBH and its
environment is an important component of the theoretical study of objects
like Sgr A*.

Besides the study of SMBH's, any astrophysical problem that involves radiative transfer and strong gravity could be tackled by general-relativistic radiative transfer codes. One example of such alternative applications is a binary black-hole system. For such systems, no analytical spacetime is known, and a numerical spacetime must be employed. Another example is the study of radiative transfer near neutron stars, for which, again, no exact metric is known (although it may be approximated by the Kerr metric, see, e.g.,~\citealt{parfrey2017}). Finally, a general-relativistic radiative transfer code could be applied to problems that do not involve compact objects, such as the propagation of radiation in an expanding FRW-spacetime.

Certain spacetimes, such as the Schwarzschild spacetime, are amenable to analytical solution of the geodesic equation (\citealt{beloborodov2002}; \citealt{defalco2016}). A semi-analytical solution to the geodesic equation in the Kerr spacetime is presented by \citealt{dexter2009}. Analytical solutions to the geodesic equation near compact objects have also been studied in the context of, for example, radiating pulsars (\citealt{poutanen2006}). Analytically or semi-analytically computed null geodesics have the important advantage of excellent spatial accuracy that is independent of integration step size. On the other hand, the analytical formulae may be expensive to evaluate, and thus a numerical code may, in some cases, offer superior performance when sensitive radiative-transfer calculations (which require a relatively small spatial integration step size) are included. Additionally, analytical codes are restricted to the set of spacetimes for which the metric and connection are known analytically.

Numerical radiative-transfer codes, capable of producing images and/or
spectra from GRMHD simulations by performing radiative-transfer
calculations in strong gravity, have been studied in a number of works
\citep{broderick2006, noble2007, dexter2009, shcherbakov2011,
  psaltis2013, dexter2016}. \cite{dexter2016}'s {\tt grtrans} is a
CPU-based code that uses a semi-analytic method to construct geodesics
and offers polarized radiative transfer in the Kerr
spacetime; \cite{ipole}'s {\tt ipole} is a numerical code that offers the same functionality, but for general spacetimes. \cite{psaltis2013}'s {\tt GRay} is a fully numerical GPU-based
{\tt CUDA} code capable of handling arbitrary spacetime metrics; it was
recently succeeded by {\tt GRay2}, a general-purpose geodesic integrator
for the Kerr spacetime \citep{gray2}. \citet{takahashi2017} have recently
presented {\tt ARTIST}, a code that is not based on a ray-tracing
algorithm, but which is capable of reproducing radiation fields (wave
fronts) in the Kerr spacetime, with the aim of including radiation
pressure in GRMHD simulations. Although both GPU and CPU codes that
perform radiative-transfer calculations are available, existing codes are
specialized to one or the other. Many are also restricted to hardware
from a single manufacturer, so that deciding which code to use may
strongly depend on the hardware available locally.

In this paper, we present a new code, 
 {\tt RAPTOR}. It was designed with two
goals in mind: minimizing the number of physical assumptions, by
supporting arbitrary spacetimes and time-dependent radiative transfer,
and maximizing flexibility of use, by supporting all commonly available
CPU and GPU hardware. 
Although the code was developed with the science case described above in mind, it may be readily applied to any astrophysical problem involving radiative transfer in strong gravity, as it is equipped to deal with numerical as well as analytical metrics. One alternative application for the code that we have explored in particular concerns visualisation of black holes in virtual reality, a project that has applications in outreach and intuitive understanding of black hole accretion \citep{vrpaper}.
Presently, we only compute the specific intensity
as seen by the observer. The next step, which involves polarized
radiative transfer, will be covered in a future paper. In this work, we
demonstrate correct operation of {\tt RAPTOR} and couple it to GRMHD
simulation data from {\tt BHAC} \citep{BHAC} and {\tt HARM2D}
\citep{gammie2003} using radiative-transfer models
\citep{moscibrodzka2009}. We also perform a detailed comparison with the
radiative-transfer code {\tt BHOSS} \citep{BHOSS} by specifying identical
initial conditions in both codes and comparing the resulting
images. Finally, in order to evaluate the accuracy of theoretical
predictions of the appearance and time-variability of SMBH accretion
flows, we apply our new code to studying the slow-light paradigm, in
which the assumption of staticity of GRMHD data is relaxed, in the
context of accreting black holes. 

The paper is organized as follows: in Section \ref{sec:geodesics}, we
present the governing equations for calculating (null) geodesics in a
curved spacetime, derive two algorithms for solving them numerically, and
test our implementations. In Section \ref{sec:rad_trans}, we discuss the
covariant radiative-transfer equation and the accompanying numerical
algorithms for solving it. Section \ref{sec:verification} contains a
library of verification and performance results for the geodesic and
radiative-transfer integrators. In Section \ref{sec:slowlight}, we apply
{\tt RAPTOR} to investigate the slow-light paradigm for imaging GRMHD
simulations.

\section{Equations of motion for light rays}\label{sec:geodesics}

The appearance and spectrum of an accreting black hole recorded by a
distant observer are strongly affected by gravitational effects such as
lensing and redshift. In order to construct an accurate image or
spectrum, these effects must be taken into account. Relativistic
ray-tracing algorithms solve the equations of motion for light in curved
spacetime, automatically taking into account all gravitational
effects. Such algorithms are discussed in this section.

\subsection{Geodesic equation}

In Newtonian physics, test particles move along straight lines in the
absence of any force. Analogously, in GR, test particles move along
so-called geodesics when in free fall, in other words, when acted upon by gravity
only. Furthermore, in GR gravitational effects on test particles are
represented through curvature of the spacetime in which the particles
move.

The structure of spacetime in GR is described by the symmetric rank-2
metric tensor $g_{\mu \nu}$ and
%
%
the motion of test particles is described by the geodesic equation
\begin{equation}\label{eqn:geodesic_eqn}
\frac{{\rm d}^2 x^{\alpha}}{{\rm d}s^2} = -\Gamma^{\alpha}_{\ \mu \nu}
\frac{{\rm d}x^{\mu}}{{\rm d}s} \frac{{\rm d}x^{\nu}}{{\rm d}s}\,.
\end{equation}
Here, $x^{\alpha}$ is the particle's position; $s$ is a scalar parameter
of the particle's world line, and $\Gamma^{\alpha}_{\ \mu \nu}$ is the
`connection' of the spacetime through which the particle
propagates. The connection depends on first derivatives of the metric and
is given by
\begin{equation}
\Gamma^{\alpha}_{\ \mu \nu} = \frac{1}{2} g^{\alpha \rho} \left[
  \partial_{\mu} g_{\nu \rho} + \partial_{\nu} g_{\mu \rho} -
  \partial_{\rho} g_{\mu \nu} \right]\,.
\label{eqn:connection_eqn}
\end{equation}
Massive particles move along `timelike' geodesics. Adopting the
Lorentzian metric signature $(-,+,+,+)$ and geometrized units, so that
$G=c=1$, the spacetime interval for massive particles of unit mass is
negative and unitary:
\begin{equation}
g_{\mu \nu} \frac{{\rm d}x^{\mu}}{{\rm d}\tau} 
\frac{{\rm d}x^{\nu}}{{\rm d}\tau} = -1\,,
\label{eqn:timelike_geodesic}
\end{equation}
where $\tau$ is the proper time measured by an observer co-moving with
the particle. Photons, which are massless and always travel at the speed
of light, travel along `null' geodesics with zero spacetime
interval:
\begin{equation}\label{eqn:null_geodesic}
g_{\mu \nu} \frac{{\rm d}x^{\mu}}{{\rm d}\lambda} \frac{{\rm
    d}x^{\nu}}{{\rm d}\lambda} = 0\,.
\end{equation}
In this case, the geodesic is parametrized by a so-called affine parameter
$\lambda$, as no proper time elapses for photons. From now on, we will
only consider null geodesics.

\subsection{Numerical integration of the geodesic equation}

In this section we present two algorithms to solve
Eq.~\eqref{eqn:null_geodesic} numerically, and discuss a number of tests of
the integrator's performance.

Since the geodesic equation ~\eqref{eqn:geodesic_eqn} represents a set of
four coupled second-order ordinary differential equations (ODE's), we
must specify both an initial position $x^{\alpha}_0$ and an initial
contravariant four-momentum vector, which in the case of radiation is the
so-called wave vector $k^{\alpha}_0$, to obtain the full geodesic
by numerical integration. As a first step, we can write
Eq.~\eqref{eqn:geodesic_eqn} as a system of eight coupled first-order
ODE's:
\begin{eqnarray}\label{eqn:eqns_to_integrate}
\frac{{\rm d} x^{\alpha}}{{\rm d}\lambda} &=& k^{\alpha}, \\ \frac{{\rm
    d} k^{\alpha}}{{\rm d}\lambda} &=& -\Gamma^{\alpha}_{\ \mu \nu}
k^{\mu} k^{\nu}.
\end{eqnarray}
We must now choose an appropriate integration scheme to solve
Eq.~\eqref{eqn:eqns_to_integrate}. Since we aim to strike a balance
between accuracy and efficiency, we present two alternatives, one of
which is more accurate while the other is less computationally expensive.

\subsubsection{Runge-Kutta integrator}

We first choose the popular 4th-order Runge-Kutta integration method
(RK4) to solve Eq.~\eqref{eqn:eqns_to_integrate} numerically. In the case
of the geodesic equation there are eight dependent variables (the
components of $x^{\alpha}$ and $k^{\alpha}$), and we must evaluate 32
`update coefficients' for the RK4 integration:
\begin{eqnarray}
C_{1,x^{\alpha}} &=& \Delta \lambda \, k^{\alpha}, \\
C_{2,x^{\alpha}} &=& \Delta \lambda \, \left( k^{\alpha} + \tfrac{1}{2} C_{1,x^{\alpha}}
\right), \\
C_{3,x^{\alpha}} &=& \Delta \lambda \, \left( k^{\alpha} + \tfrac{1}{2} C_{2,x^{\alpha}}
\right), \\
C_{4,x^{\alpha}} &=& \Delta \lambda \, \left( k^{\alpha} + C_{3,x^{\alpha}} \right), \\
C_{1,k^{\alpha}} &=& \Delta \lambda \, f^{\alpha}\left( \lambda, x^{i},
k^{i} \right), \\
C_{2,k^{\alpha}} &=& \Delta \lambda \, f^{\alpha}\left( \lambda + \tfrac{1}{2} \Delta \lambda,
x^{i} + \tfrac{1}{2} C_{1,x^{i}}, k^{i} +
\tfrac{1}{2} C_{1,k^{i}}\right), \\
C_{3,k^{\alpha}} &=& \Delta \lambda \, f^{\alpha}\left( \lambda + \tfrac{1}{2} \Delta \lambda,
x^{i} + \tfrac{1}{2} C_{2,x^{i}}, k^{i} +
\tfrac{1}{2} C_{2,k^{i}}\right), \\
C_{4,k^{\alpha}} &=& \Delta \lambda \, f^{\alpha}\left( \lambda + \Delta \lambda, x^{i}
+ C_{3,x^{i}}, k^{i} + C_{3,k^{i}}\right),
\label{eqn:RK4_geodesic}
\end{eqnarray}
where $\Delta \lambda$ is the discrete increment in the affine parameter,
$f^{\alpha}$ represents the right-hand side of
Eq.~\eqref{eqn:geodesic_eqn}, and $i$ is a shorthand indicating that all
components of $x^{\alpha}$ and $k^{\alpha}$ appear as variables of
$f^{\alpha}$, that is, $f^{\alpha}\left( \lambda, x^{i}, k^{i}
\right)=f^{\alpha}\left( \lambda, x^{1}, x^{2}, x^{3},
x^{4},k^{1},k^{2},k^{3},k^{4} \right)$. Having calculated all update
coefficients using Eq.~\eqref{eqn:RK4_geodesic}, we can compute the new
values for $x^{\alpha}$ and $k^{\alpha}$ as 
\begin{eqnarray}
x^{\alpha}_{new} &=& x^{\alpha} + \tfrac{1}{6} \left( C_{1,x^{\alpha}} + 2
C_{2,x^{\alpha}} + 2 C_{3,x^{\alpha}} + C_{4,x^{\alpha}} \right) + O\left({\Delta
\lambda}^5\right)\,, \nonumber \\ \\
k^{\alpha}_{new} &=& k^{\alpha} + \tfrac{1}{6} \left( C_{1,k^{\alpha}} + 2
C_{2,k^{\alpha}} + 2 C_{3,k^{\alpha}} + C_{4,k^{\alpha}} \right) + O\left({\Delta
\lambda}^5\right)\,. \nonumber \\ 
\label{eqn:RK4_geodesic_update}
\end{eqnarray}

\subsubsection{Verlet integrator}

Although the RK4 integrator is accurate, more efficient integrators
exist. Evaluating the connection coefficients is the most computationally
expensive operation in our geodesic integration algorithms; structuring
our code in this manner allows us to maintain a general scheme that is
capable of interfacing with GRMHD simulations in different coordinate
systems, as well as of handling spacetimes that are completely arbitrary
such as those in the framework of \citet{Rezzolla2014} and
\citet{Konoplya2016a}, and recently employed by
\citet{Younsi2016}. \cite{dolence2009} have presented the velocity Verlet
algorithm \citep{velocityverlet} as a more efficient alternative to
the RK4 algorithm, as it relies on fewer evaluations of the connection
coefficients:
\begin{eqnarray}
x^{\alpha}_{n+1}&=&x^{\alpha}_n + k^{\alpha}_n \Delta \lambda + \frac{1}{2} \left( \frac{{\rm d} k^{\alpha}}{{\rm d}\lambda}\right)_n \left( \Delta \lambda \right)^2 \label{subeq1}\,, \\
k^{\alpha}_{n+1,p}&=&k^{\alpha}_n+\left( \frac{{\rm d} k^{\alpha}}{{\rm d}\lambda}\right)_n \Delta \lambda\,, \label{subeq2}, \\
\left( \frac{{\rm d} k^{\alpha}}{{\rm d}\lambda}\right)_{n+1}&=&-\Gamma^{\alpha}_{\ \mu \nu} \left( x^{\alpha}_{n+1} \right) k^{\mu}_{n+1,p} \ k^{\nu}_{n+1,p}\,, \label{subeq3} \\
k^{\alpha}_{n+1}&=&k^{\alpha}_n+\frac{1}{2} \left[ \left( \frac{{\rm d}
    k^{\alpha}}{{\rm d}\lambda}\right)_n + \left( \frac{{\rm d}
    k^{\alpha}}{{\rm d}\lambda}\right)_{n+1} \right] \Delta \lambda \,. \label{subeq4}
\label{eqn:Verlet_geodesic_update}
\end{eqnarray}
The accuracy of this algorithm can be improved by using the result of
Eq.~\eqref{subeq4} to recompute the derivative (Eq.~\eqref{subeq3}) with
$k^{\mu}_{n+1,p}=k^{\mu}_{n+1}$, and then reevaluating
Eq.~\eqref{subeq4}.

In this paper we have restricted our investigations to the Kerr spacetime,
which represents an uncharged black hole with (in the general case)
non-zero angular momentum characterized by the {spin parameter}
$a:=J/M^2$, where $J$ is the black-hole's angular momentum and $M$ its
mass. Near the black-hole event horizon, as well as near the symmetry
axis (the spin axis), numerical integration becomes difficult due to
coordinate singularities. The following adaptive step-size routine
introduced by \citet{noble2007} and \citet{dolence2009} is adopted for
accuracy and efficiency in difficult regions by reducing the step-size
$\Delta \lambda$ in these cases:
\begin{eqnarray}
{\rm d}\lambda&=&\frac{1}{{\left| {\rm d}\lambda_{x^1} \right|^{-1}} +
  {\left| {\rm d}\lambda_{x^2} \right|^{-1}} + {\left| {\rm
      d}\lambda_{x^3} \right|^{-1}}}\,,
\label{adaptivestep_1}
\end{eqnarray}
where
\begin{eqnarray}
{\rm d}\lambda_{x^1} &:=& \epsilon \ / \left( \left| k^r \right| + \delta
\right)\,, \\ {\rm d}\lambda_{x^2} &:=& \epsilon \ \text{min}
\left(x^{\theta}, 1 - x^{\theta} \right) / \left( \left| k^{\theta}
\right| + \delta \right)\,, \\ {\rm d}\lambda_{x^3} &:=& \epsilon \ /
\left( \left| k^{\phi} \right| + \delta \right)\,.
\label{adaptivestep_2}
\end{eqnarray}
Here, $\delta$ is a very small (positive) number that protects against
dividing by 0, while $\epsilon$ is a (positive) scaling parameter by
which one can influence the scale of all steps.

\subsection{Initial conditions: the virtual camera}

Specific to the case of a Kerr black hole, consider a distant observer
whose inclination angle with respect to the black-hole rotation axis is
given by $i$ so that an inclination angle of 90\degr means the observer
is in the black-hole equatorial plane, while an inclination angle of
0\degr means that we are looking at the black hole along the direction of
its spin axis. In the observer's image plane, which is oriented so that
its vertical axis is aligned with the black-hole spin axis (for any
non-zero inclination angle), we define the impact parameters
$\alpha$ (the distance from the black-hole rotation axis) and $\beta$
(the distance in the direction perpendicular to $\alpha$). The wave
vector $k^{\alpha}$ is then constructed following \citet{bardeen1972b} as
\begin{eqnarray}
L&=& -\alpha E \sqrt{1-\cos^2{i} }\,, \\
Q&=& E^2 \left[ \beta^2 + \cos^2{i} \left(\alpha^2-1\right) \right]\,,  \\
k_t&=&-E\,, \\
k_{\phi}&=&L\,, \\
k_{\theta}&=&\text{sign}(\beta)\, \sqrt{\left| Q-L^2 \cot^2{\theta} + E^2 \cos^2{\theta} \right|}\,,
\label{eqn:constructingK}
\end{eqnarray}
where $E$ is the wave vector's total energy, $L$ is the projection of the
angular momentum parallel to the black- hole spin axis, and $Q$ is the
Carter constant \citep{carter1968}. The radial component of the wave
vector, $k^r$, is fixed by demanding that it is a null vector, that is,
$k_{\alpha}k^{\alpha}=0$ ($k^r$ may be positive or negative,
corresponding to out- and ingoing rays, respectively). Boyer-Lindquist
(BL) coordinates \citep{boyer1967} are used to construct the initial wave
vector.

\subsection{Coordinate systems and transformations}

One of the main purposes of {\tt RAPTOR} is to integrate the
radiative-transfer problems generic black-hole spacetimes such as those
proposed in alternative to GR theories of gravity
\citep{Rezzolla2014}. Hence, null-geodesic integration and the
radiative-transfer equations described in the next section are formulated
in a way that is independent of the choice of coordinate system or of the
geometry of the spacetime. In view of this, {\tt RAPTOR} has been
constructed so as to switch easily among various grids and geometries. In
particular, when considering the solution of the radiative-transfer
equation near black holes, it is important that the coordinate system is
accurate near the horizon, and compatible with GRMHD simulations; the
latter customarily adopt a spherical-polar coordinate system using a
logarithmic scale for the radial coordinate and a denser polar coordinate
mapping near the equatorial plane (so-called modified coordinate systems,
see, e.g., \citealt{gammie2003}). In this paper, we utilized four different
coordinate systems for the Kerr spacetime, namely: the aforementioned
Boyer-Lindquist coordinates, the modified Boyer-Lindquist (MBL)
coordinates, the logarithmic Kerr-Schild (KS) coordinates
\citep{kerr1965}, and the modified Kerr-Schild (MKS) coordinates
\citep{gammie2003}. Some of the transformation laws between these
coordinate systems are given explicitly in Appendix \ref{appB}, while a
test of the code performance is presented in Appendix
\ref{sec:code_performance}.

\section{Radiative transfer}\label{sec:rad_trans}

Having previously computed the relevant null geodesics, an algorithm is
needed that will perform radiative-transfer calculations along them. In
this section we introduce the relevant equations. Our present algorithm
does not include radiation refraction effects due to the plasma (which is
a good approximation if the radiation frequency is greater than the
plasma frequency, $\nu_p=8980\, \ n_{\rm e}^{1/2}$, where $n_{\rm e}$ is
the electron number density), all forms of scattering, and polarization although
all of these effects can be incorporated in a ray-tracing simulations
(see, e.g., \citealt{broderick2006}, \citealt{dolence2009},
or~\citealt{dexter2016}). However, {\tt RAPTOR} integrates the radiative
transfer equations taking into account changes in the plasma structure
during the light transport (Sect.~\ref{sec:slowlight}), which is usually
neglected. Hence, our code is suitable for first-principles study of time
variability of mock observations of accreting black holes or any other
compact objects surrounded by plasma.

\subsection{Covariant radiative-transfer equation}

The transfer equation for the Lorentz invariant quantity
$I_{\nu}/{\nu^3}$, where $I_{\nu}$ is the specific intensity of radiation
at frequency $\nu$, is given by \citep{lindquist1966}:
\begin{equation}
\frac{{\rm d} }{{\rm d}\lambda} \left( \frac{I_{\nu}}{\nu^3} \right) =
\frac{j_{\nu}}{\nu^2} - \nu \alpha_{\nu}
\left(\frac{I_{\nu}}{\nu^3}\right)\,,
\label{eqn:radtranseqn}
\end{equation}
where again $\lambda$ is the affine parameter, which we defined to be
increasing as the ray travels from the plasma toward the observer,
$\nu$ is the photon's frequency, $j_{\nu}$ is the plasma emission
coefficient, and $\alpha_{\nu}$ is the plasma absorptivity. We note that all
of these physical quantities are computed in an inertial frame that is
co-moving with the plasma (hereafter the fluid frame). In the present
case of unpolarized radiative transfer, the necessary transformation
between frames can be achieved simply by computing the ray's frequency in
the fluid frame:
\begin{equation}
\nu=-k^{\alpha} u_{\alpha}\,,
\label{eqn:normalizing}
\end{equation}
and using $\nu$ to relate the fluid frame emission and absorption
coefficients to their Lorentz-invariant counterparts, so that we do not
have to construct the fluid frame explicitly, as is the case with
polarized radiative transfer.

The intensity seen by the observer, $I_{\nu,\rm{obs}} =
I_{\nu}\left(\lambda_{\rm obs}\right)$, is then obtained by integrating
Eq.~\eqref{eqn:radtranseqn} from $\lambda_0$ to $\lambda_{\rm obs}$ and
subsequently converting from the Lorentz-invariant quantity
${I_{\nu}}/{\nu^3}$ to the specific intensity in the observer frame
using
\begin{equation}
I_{\nu,\rm{obs}} = \frac{I_{\nu}}{\nu^3} \nu_{\rm{obs}}^3\,.
\label{eqn:invariant_to_intensity}
\end{equation}

Although it is possible to integrate Eq.~\eqref{eqn:radtranseqn}
directly, more accurate numerical results can be obtained by employing
the formal solution to the transfer equation (see e.g.,
\cite{dexter2009}), which recasts $I_{\nu}$ as a function of the optical
depth $\tau_{\nu}$ given by:
\begin{equation}
\tau_{\nu} \left(\lambda\right)=\int_{\lambda_0}^{\lambda} \nu
\alpha_{\nu}{\rm d}\lambda'\,.
\label{eqn:opticaldepth}
\end{equation}
We note that the optical depth, which describes the fraction of photons that
pass through a certain absorbing volume, is a Lorentz-invariant
quantity. The formal solution to the radiative-transfer equation is then
\begin{equation}
\frac{I_{\nu}}{\nu^3} \left( \tau_{\nu} \right) = \frac{I_\nu}{\nu^3}
\left( \tau_{\nu,0} \right) \exp{\left(-\tau_{\nu}\right)}+
\int_{\tau_{\nu,0}}^{\tau_{\nu}} \exp{-\left( \tau_{\nu} -\tau_{\nu}'
  \right)} \frac{j_{\nu}}{\nu^3 \alpha_{\nu}} {\rm d}\tau_{\nu}'\,,
\label{eqn:formalsoln}
\end{equation}
where $\tau_{\nu,0}$ defines the location along the null geodesic where
integration begins. We can repeatedly solve Eq.~\eqref{eqn:formalsoln} for
each integration step. Assuming that $j_{\nu}$ is constant over the step,
we obtain
\begin{equation}
\frac{I_{\nu}}{\nu^3} \left( \tau_{\nu} \right) = \frac{I_\nu}{\nu^3}
\left( \tau_{\nu,0} \right) \exp{\left(-\tau_{\nu}\right)}+
\frac{j_{\nu}}{\nu^3 \alpha_{\nu}} \left( 1 -
\exp{\left(-\tau_{\nu}\right)} \right)\,.
\label{eqn:formalsoln2}
\end{equation}

The reason for which using Eq.~\eqref{eqn:formalsoln2} yields better
numerical results than direct integration of Eq.~\eqref{eqn:radtranseqn}
with, for example, an explicit Euler scheme, is that in the latter case, the
intensity may become negative if we take too large an integration step
whenever ${j_{\nu}}/{\nu^2}<\nu \alpha_{\nu} \left( {I_{\nu}}/{\nu^3}
\right)$, thus producing grossly incorrect results. Equation
\eqref{eqn:formalsoln}, on the other hand, contains an integrand that is
always positive.

\begin{figure}
\includegraphics[width=0.475\textwidth]{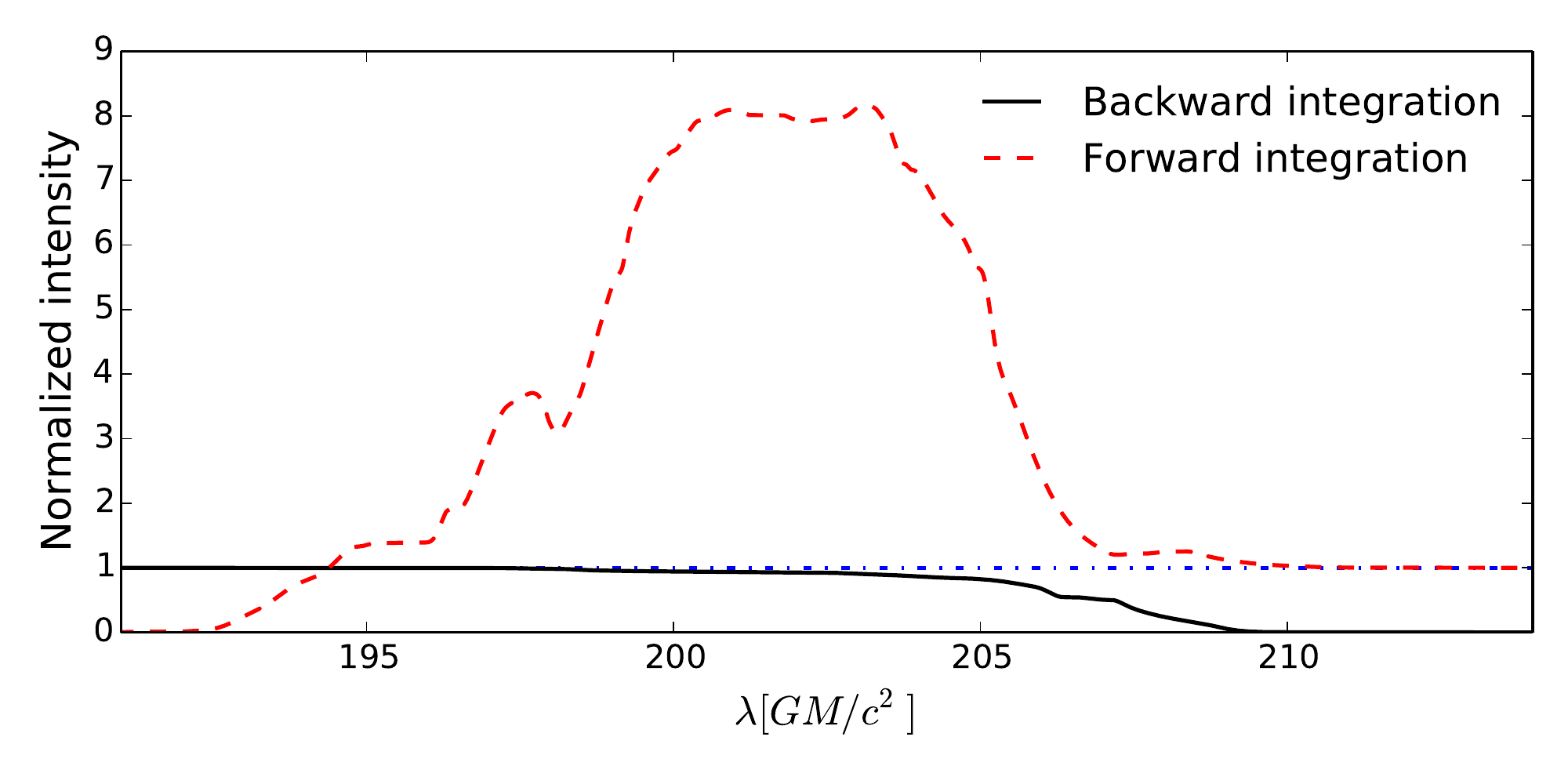}
\includegraphics[width=0.475\textwidth]{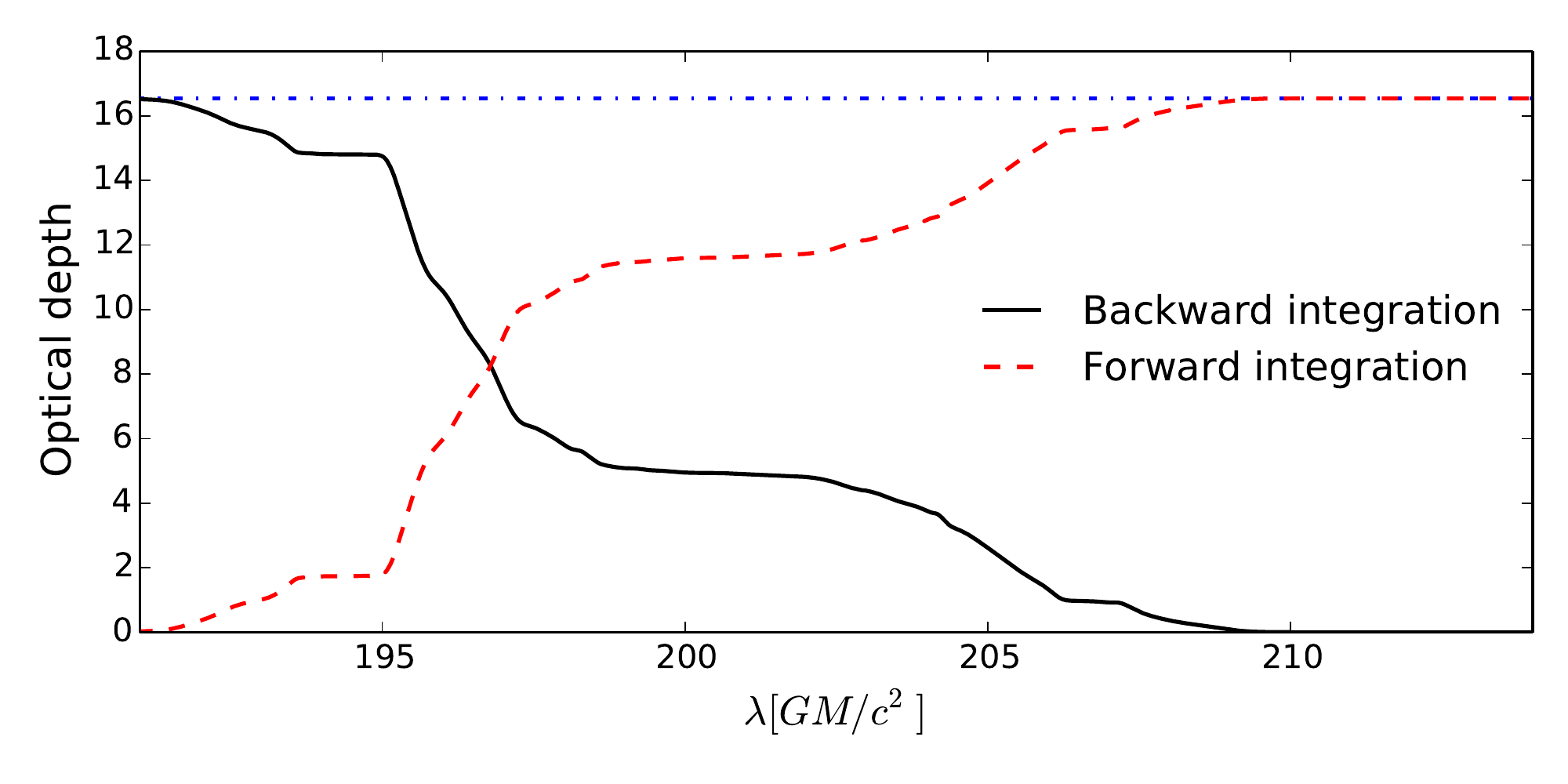}
\caption{Plots of radiative-transfer variables along a particular null
  geodesic that intersects a radiating plasma near a Kerr black hole (see
  Fig.~\ref{fig:backward_vs_forward_ray}). Top: plots of the local
  specific intensity $I_{\nu}$ (dashed red curve), which is computed by
  integrating Eq.~\eqref{eqn:formalsoln} in the direction of increasing
  affine parameter $\lambda$, and of the specific intensity seen by the
  observer, $I_{\nu,\rm{obs}}$ (solid black curve), which is computed by
  integrating Eq.~\eqref{eqn:backward_transfer} in the direction of
  decreasing $\lambda$. The intensity is normalized with respect to the
  observed intensity in both cases, so that we expect both curves to
  approach unity with continued integration, which is illustrated using
  the blue guideline. We note that $I_{\nu}$ is a non-monotonic function of
  $\lambda$ while $I_{\nu,\rm{obs}}$ is a monotonic function of
  $\lambda$. We also note that for $I_{\nu,\rm{obs}}$ and $\tau_{\nu,\rm{obs}}$, integration proceeds from right to left in these plots, and that $I_{\nu,\rm{obs}}$ converges much faster than $I_{\nu}$. Bottom: plots of the optical depth for the same ray. }\label{fig:backward_vs_forward}
\end{figure}

\begin{figure}
\includegraphics[width=0.475\textwidth]{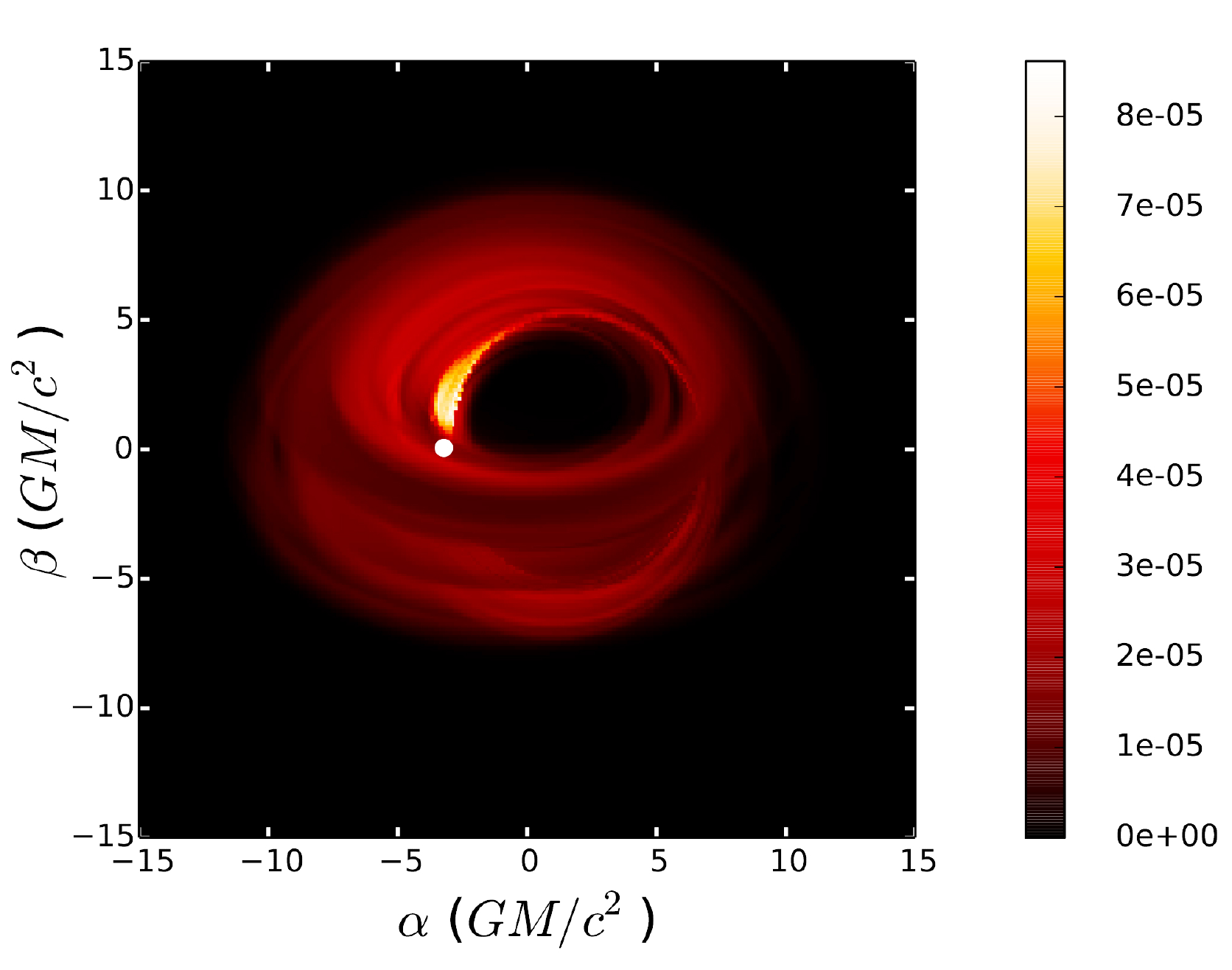}
\caption{Synchrotron emission map created using a GRMHD simulation (see
  Sec.~\ref{sec:grmhd_convergence}) made using the {\tt HARM} code
  \citep{gammie2003}. Intensity is given in units of Jy pixel$^{-2}$. The
  observer frequency is 100 GHz, the inclination angle is 30 degrees, and
  the disk-dominated emission model discussed in Sec.~\ref{sec:slowlight}
  is used. Note the optically-thick accretion disk and strong emission to
  the left of the black-hole shadow (this region appears bright due to
  relativistic beaming). The ray investigated in
  Fig.~\ref{fig:backward_vs_forward} is marked by a white dot in this
  image. This particular ray, travelling from the plasma to the observer,
  first passes through the bright, relativistically-boosted region, and
  then through an optically-thick, non-boosted region of the disk,
  suggesting that the specific intensity along this ray will sharply
  increase, then decrease, before reaching the observer. Figure
  \ref{fig:backward_vs_forward} shows that this is indeed the
  case. }\label{fig:backward_vs_forward_ray}
\end{figure}

An even more numerically advantageous method for calculating the
intensity seen by the observer can be obtained by integrating
$I_{\nu,\rm{obs}}$ in the opposite direction along the null geodesic with
respect to the previous schemes, in other words, from the observer toward the
plasma, by splitting Eq.~\eqref{eqn:formalsoln2} into separate equations
for $I_{\nu}$ and $\tau_{\nu}$ \citep{younsi2012}. In fact, in this
direction, $I_{\nu,\rm{obs}}$ must be a monotonically increasing function
of the affine parameter $\bar{\lambda}$, which we defined to increase in
the opposite spatial direction of the previously used affine
parameter $\lambda$ ($\bar{\lambda}$ increases as the ray travels further
away from the observer). This is because the specific intensity of the
ray seen by the observer can only increase, never decrease, by
continued integration further away from the camera (see
Figs.~\ref{fig:backward_vs_forward} and \ref{fig:backward_vs_forward_ray}
for an illustration of the difference between the two strategies).

This method offers three additional advantages over the two methods
described above. Firstly, it allows simultaneous integration of the null
geodesic and of the specific intensity, as well as cutting off the
null-geodesic integration when the optical depth of the material between
the current location and the camera becomes higher than a threshold value
(any radiation emitted beyond this point will be severely attenuated and
thus be negligible for the observer). Secondly, it relieves the need for
a data structure to store the geodesic in memory before performing
radiative-transfer calculations (since we simply integrate
$I_{\nu,\rm{obs}}$ along with the geodesic itself). Thirdly, with this
method, it is possible to compute appropriate integration step-sizes for
both the null geodesic and the radiative-transfer integration, and then
pick the minimum of the two. This avoids situations where the
null-geodesic integration, performed separately as if in a vacuum, takes
a large step through a plasma that is optically thick at a range of
frequencies of interest, yielding inaccurate radiative-transfer
calculations.

An expression for $I_{\nu,\rm{obs}}$ can be constructed by considering a
ray travelling backward from the camera into a radiating plasma. At each
point along the ray's path, we computed the optical depth of the material
between the current location, $\lambda$, and the observer's location,
$\lambda_{\rm obs}$ as
\begin{equation}
\tau_{\nu,\rm{obs}}\left(\lambda\right) = \int^
{\lambda} _ {\lambda_{\rm obs}} \alpha_\nu (\lambda') \ \nu \ {\rm d}\lambda'\,.
\end{equation}
We can then compute the local invariant emission coefficient,
$\left({j_\nu}/{\nu^2}\right)$, and scale it by
$\exp{\left(-\tau_{\rm{obs}}\right)}$, resulting in the following
expression for the contribution of point $\lambda$ on the geodesic to the
total observed intensity:
\begin{equation}
\frac{{\rm d}}{{\rm d}\bar{\lambda}}\left(
\frac{I_{\nu,\rm{obs}}}{\nu_{\rm obs}^3} \right) = 
\frac{j_{\nu}}{\nu^2}
\exp{\left(-\tau_{\nu,\rm{obs}}\left(\bar{\lambda}\right)\right)}\,.
\label{eqn:backward_transfer}
\end{equation}
Integrating over the geodesic then yields the total invariant intensity
at the observer, ${I_{\nu,\rm{obs}}}/{\nu_{\rm obs}^3}$, which is
converted to $I_{\nu,\rm{obs}}$ using
Eq.~\eqref{eqn:invariant_to_intensity} as before.

We have implemented all three integration strategies discussed in this
section in {\tt RAPTOR} and found that integration of
Eq.~\eqref{eqn:backward_transfer} produces the most accurate results in
the least amount of time.

\begin{figure*}
\centering
\begin{subfigure}[b]{0.245\textwidth}
	\includegraphics[width=\textwidth]{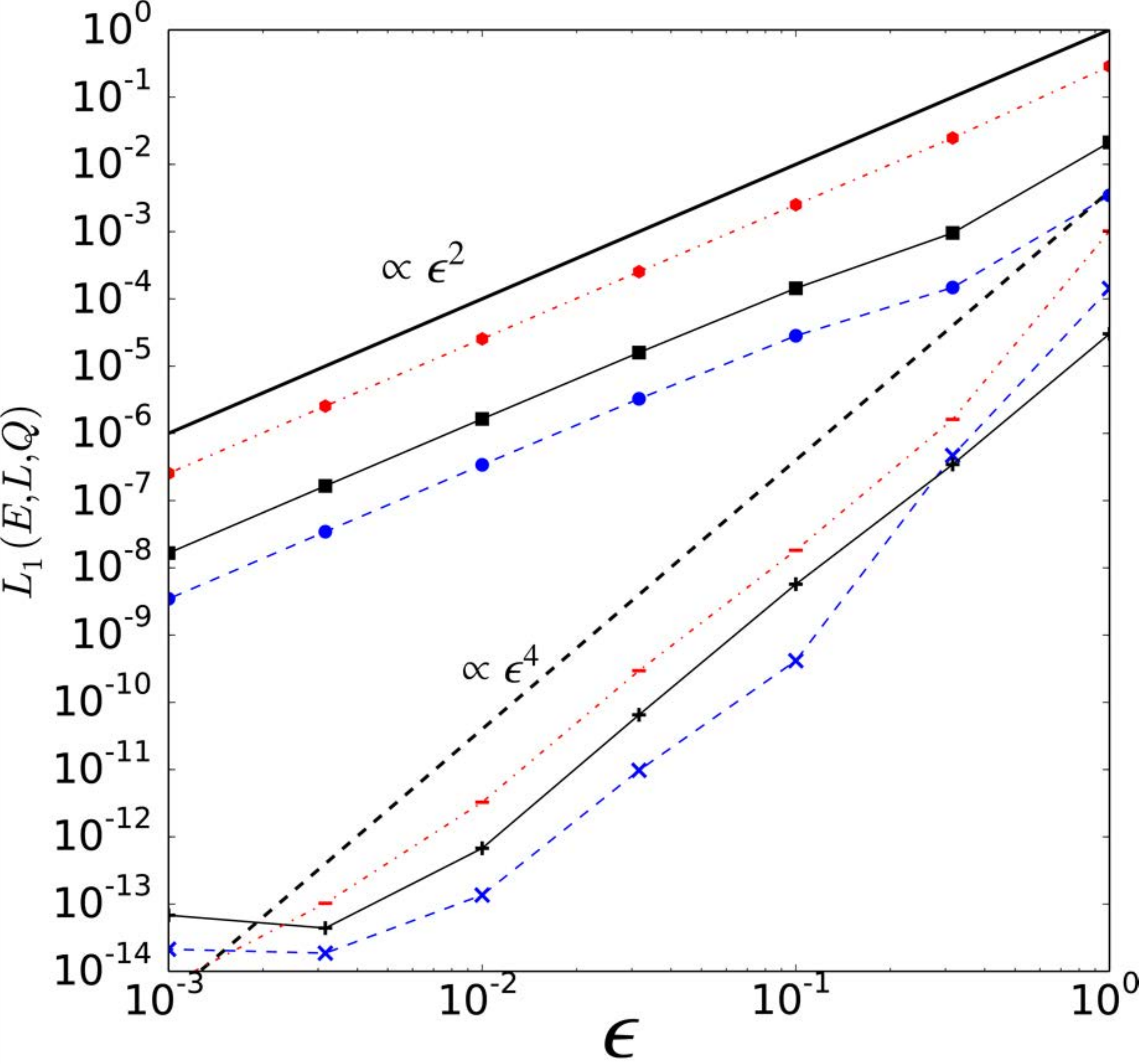}
	\caption{BL coordinates.}
\end{subfigure}
\begin{subfigure}[b]{0.245\textwidth}
	\includegraphics[width=\textwidth]{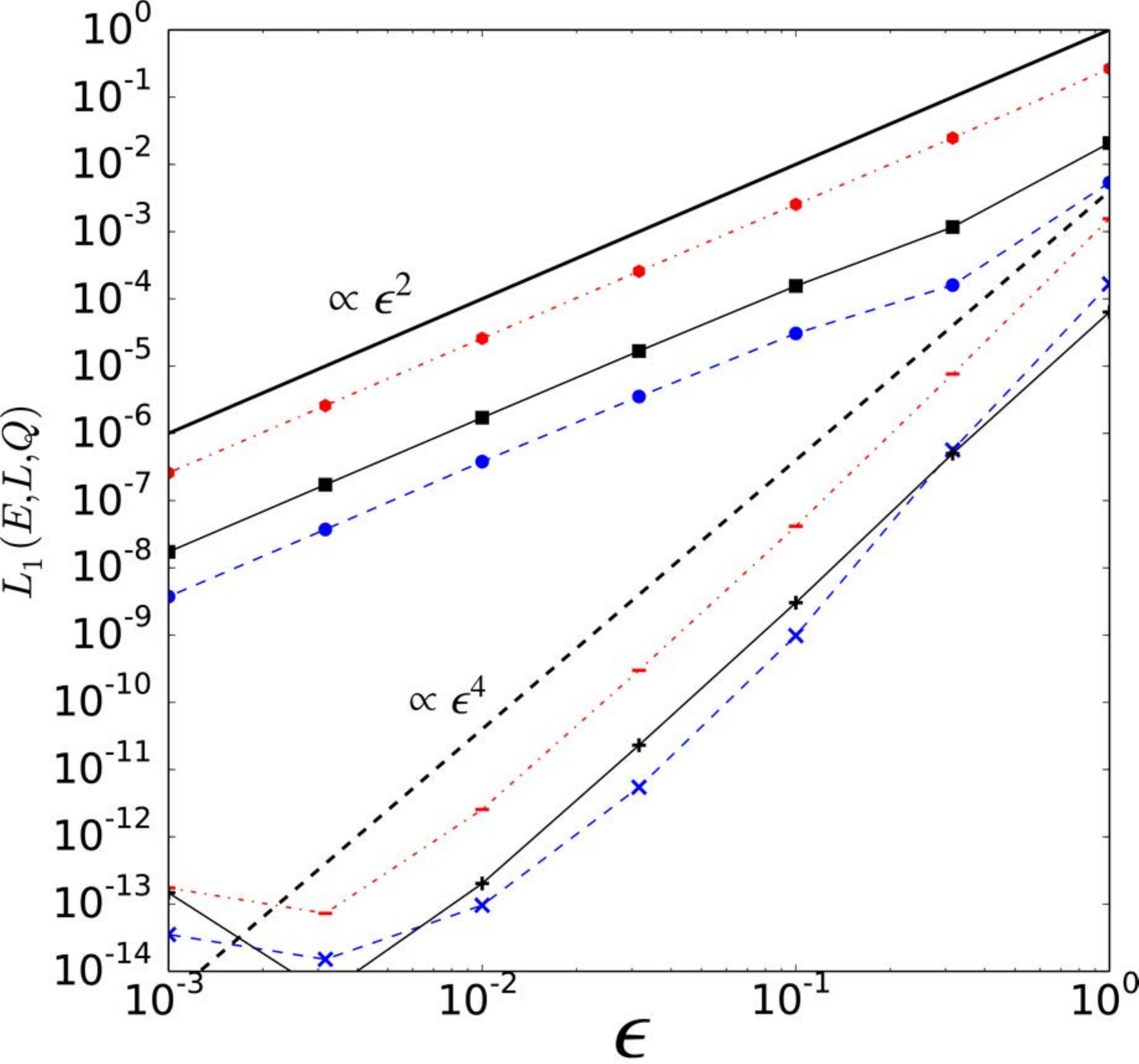}
	\caption{BL coordinates.}
\end{subfigure}
\begin{subfigure}[b]{0.245\textwidth}
	\includegraphics[width=\textwidth]{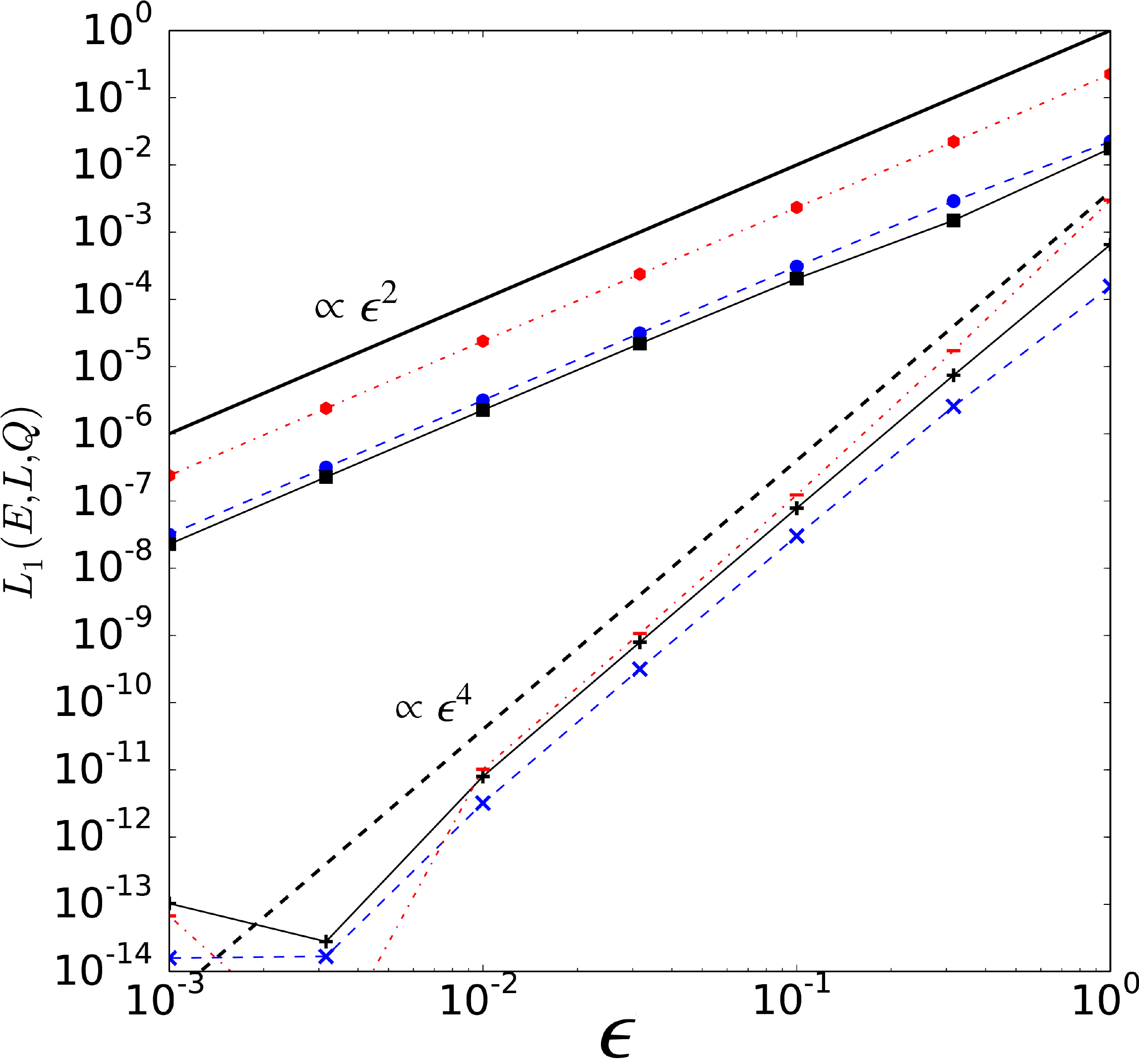}
	\caption{KS coordinates.}
\end{subfigure}
\begin{subfigure}[b]{0.245\textwidth}
	\includegraphics[width=\textwidth]{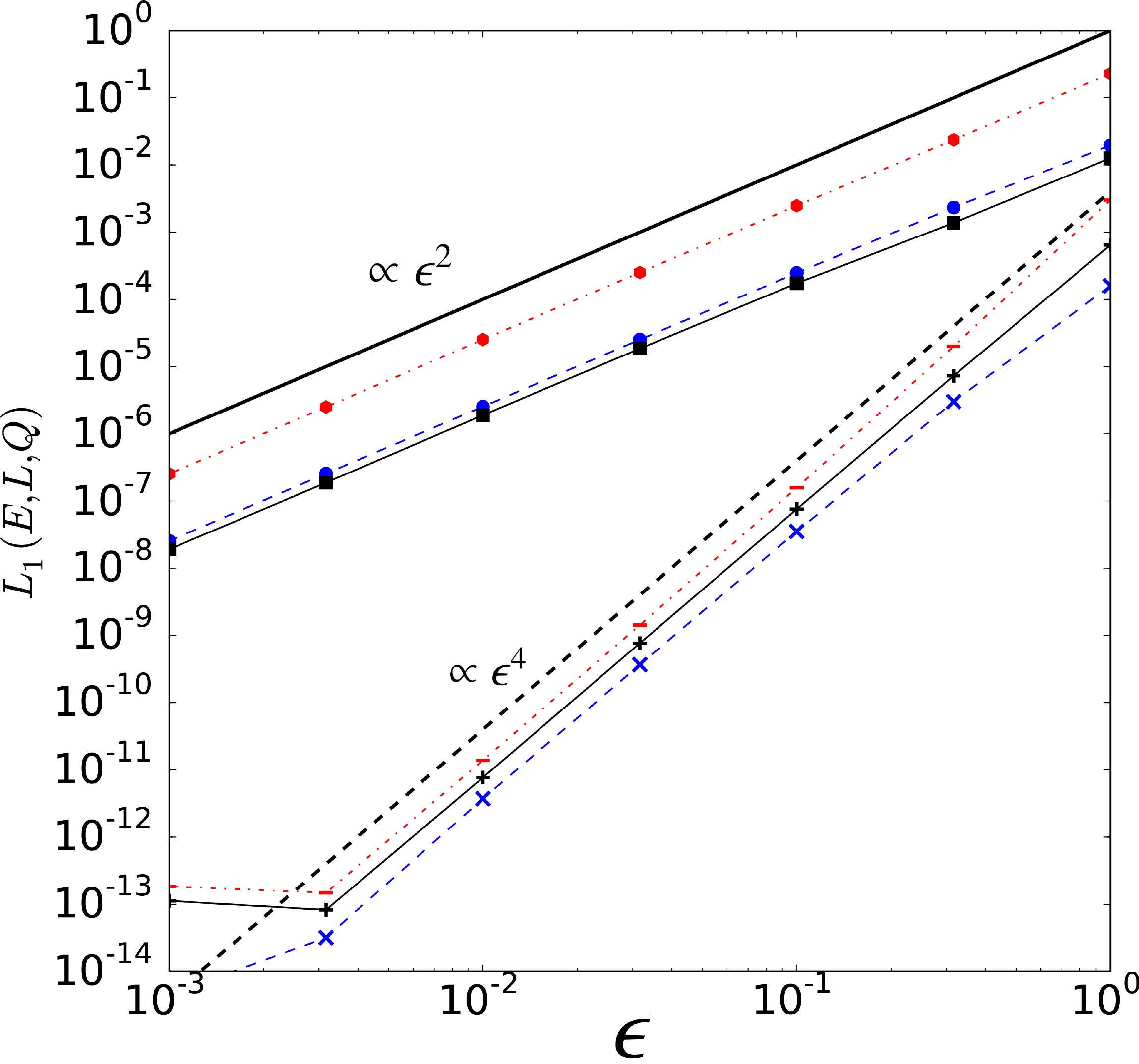}
	\caption{MKS coordinates.}
\end{subfigure}
\caption{Convergence plots for the $L_1$-error norms of the conserved
  quantities $E$ (dashed blue), $L$ (solid black), and $Q$ (dotted red)
  for the orbit with parameters $i=90\degr$, $\alpha = -3.22$, $\beta =
  1.12$, $a = 0.998$ with different integrator settings. The errors are
  computed at a fixed coordinate time $t_{\rm{final}}$. Two guide lines
  with slopes -2 (solid) and -4 (dashed) are drawn. The $+$, $-$, and
  $\times$-symbols represent respectively $L$, $Q$, and $E$, for the RK4
  integrator. The square, hexagon, and circle symbols represent the same
  quantities for the Verlet integrator. }\label{fig:convergence}
\end{figure*}

\section{Code verification}
\label{sec:verification}

Since the program consists of two parts, namely the integration of the
geodesics and the integration of the radiative-transfer equation, correct
results must be verified for both. In Section
\ref{section:geodesicverification} we verify the correct integration of
null geodesics, while in Section \ref{section:radtransverification} we
focus on the radiative-transfer computations.

\subsection{Verification of geodesic integration}
\label{section:geodesicverification}


In his seminal work, \citet{carter1968} presented three conserved
quantities associated with the orbits of photons (or particles) in the
Kerr spacetime:
\begin{eqnarray}
E&=&-k_t,\\ L&=&k_{\phi}, \\ Q&=& k_{\theta}^2 + \cos^2{\theta} \left[
  a^2 \left( \mu^2 - k_t^2 \right) + k_{\phi}^2 / \sin^2{\theta}
  \right]\,,
\label{eqn:constants_of_motion}
\end{eqnarray}
where $Q$ is the so-called Carter constant. Figure
\ref{fig:convergence} shows, in terms of the $L_1$-error norms, that
these quantities are conserved as expected and that the errors converge
respectively to second and fourth order in step-size for the Verlet and
RK4 algorithms, for all coordinate systems.

\begin{figure*}
\centering
\begin{subfigure}[b]{0.475\textwidth}
	\includegraphics[width=\textwidth]{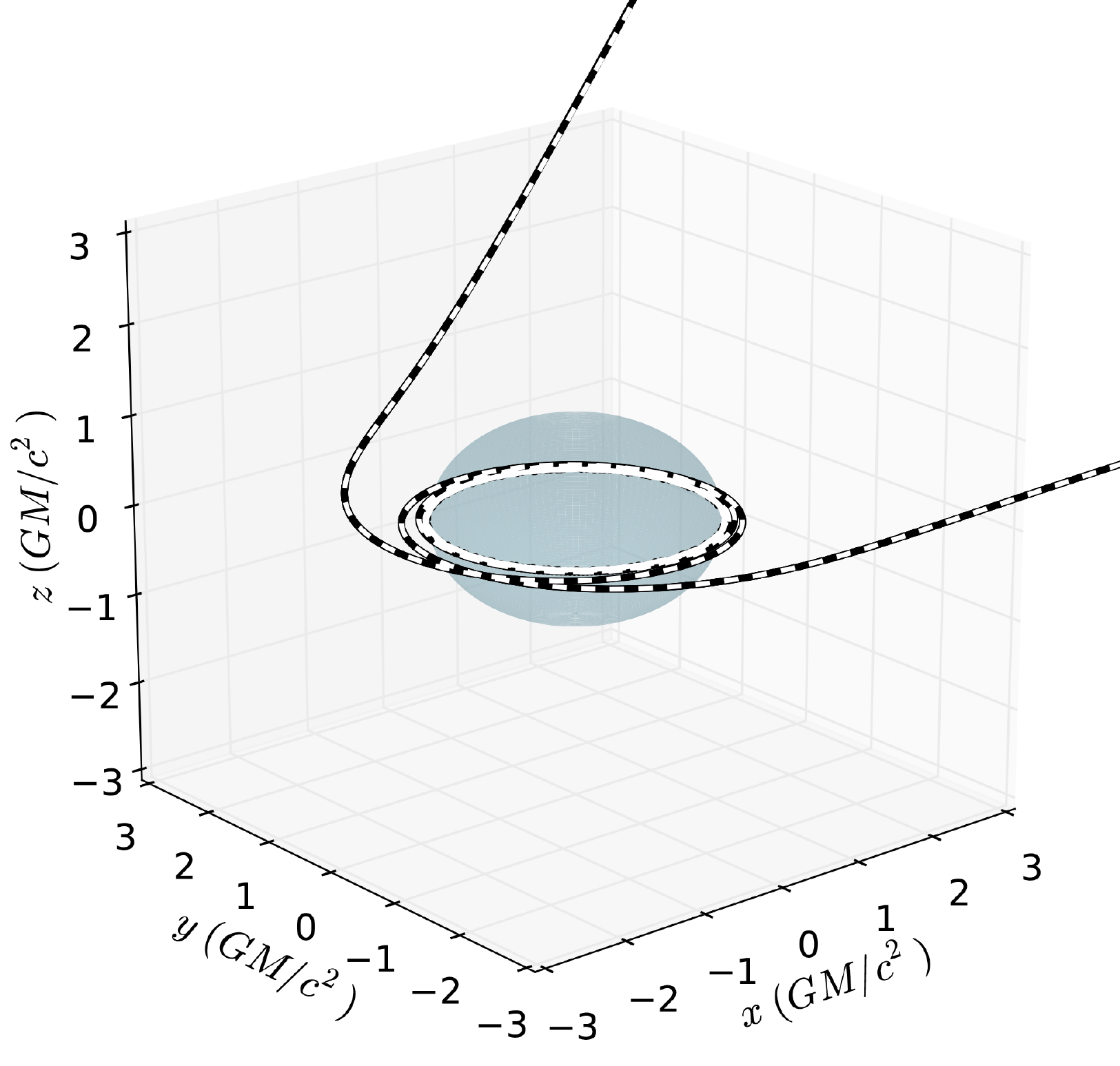} 
	\caption{Orbit 1.}
\end{subfigure}
\begin{subfigure}[b]{0.475\textwidth}
	\includegraphics[width=\textwidth]{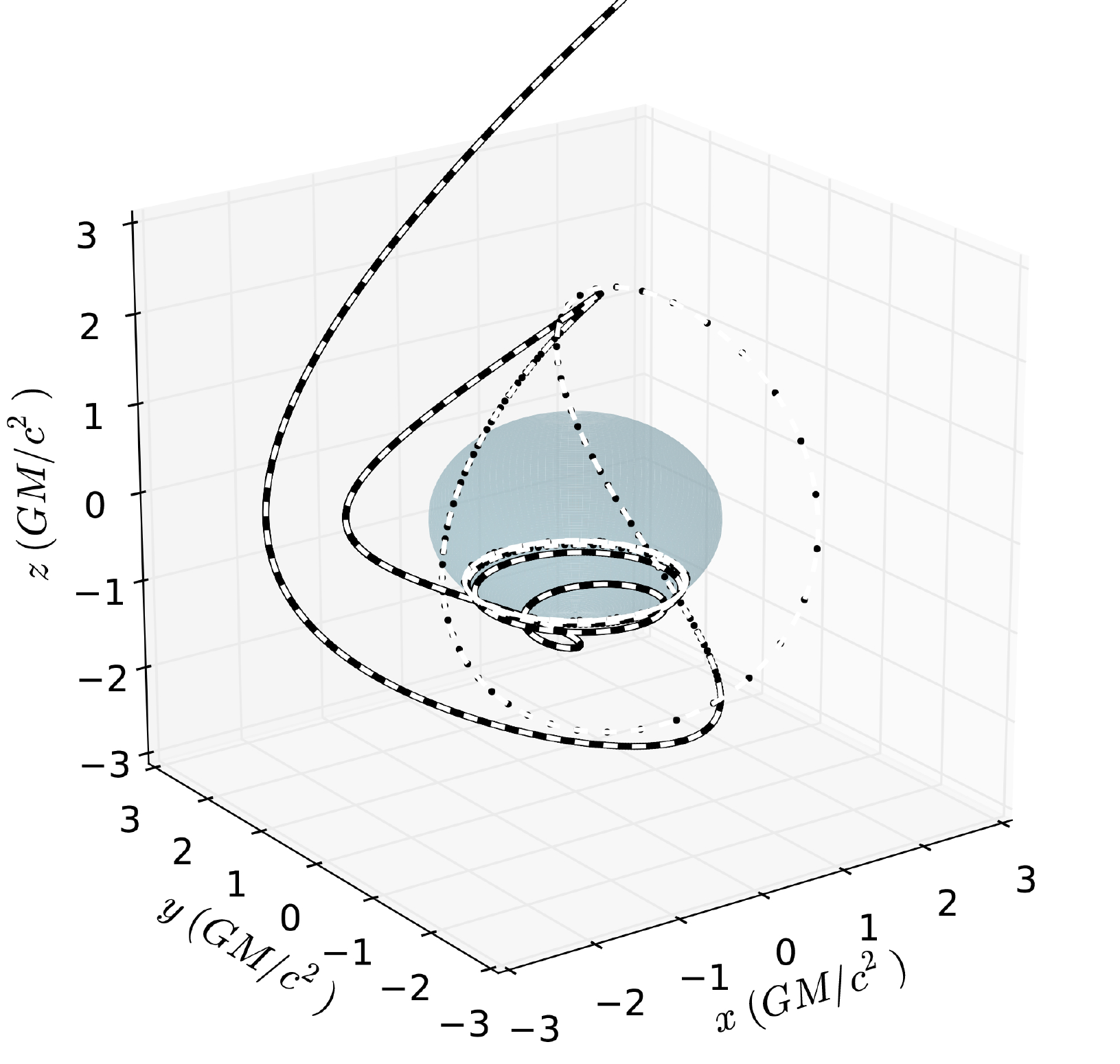} 
	\caption{Orbit 2. }
\end{subfigure}
\caption{Comparison of null geodesics computed by the semi-analytical
  integrator {\tt geokerr} to those computed by our fully numerical
  code. In each case, the geodesic represented by the black dots was
  computed using {\tt geokerr} while the geodesic represented by the
  white dotted line is the output of {\tt RAPTOR} (dots were chosen for
  the former geodesic because its variable step-size can otherwise create
  interpolation issues). The gray spheroid represents the black hole's
  outer event horizon. MKS coordinates were used with
  $\epsilon=0.001$. The parameters for these geodesics are listed in
  Table \ref{table:firsttable}. }\label{fig:diff_geo}
\end{figure*}

Another interesting test for the geodesic integration comes from
considering those null geodesics in the Kerr spacetime that have a
complex morphology, looping around the horizon many times before plunging
in or escaping. Such pathological geodesics are a good test for the
code's performance in a worst-case scenario because even small errors
lead to deviations (e.g., absorption into the event horizon). Table
\ref{table:firsttable} shows the parameters for two such geodesics.
\begin{table}
\centering
\begin{tabular}{l l l}
\hline
\hline
 parameter & Orbit 1 & Orbit 2 \\
\hline
$a$ & 0.998 & 0.998 \\
$i$ & $90\degr$ & $60\degr$ \\
$\alpha$ & -2.1109 & -0.001 \\
$\beta$ & 0 & 4.752\\
\hline
\end{tabular}
\caption{Parameters for the two `difficult' null geodesics under
  investigation.}~\label{table:firsttable}
\end{table}
Since {\tt geokerr} is a semi-analytical code, it provides an excellent
reference for the accuracy of a numerical scheme. Plots of the null
geodesics described in Table \ref{table:firsttable}, produced by both
{\tt geokerr} and {\tt RAPTOR}, are shown in Fig.~\ref{fig:diff_geo}.

\subsection{Verification of radiative-transfer calculations}~\label{section:radtransverification}

\subsubsection{Emission line profiles and images of a thin accretion disk}

\begin{figure}
\includegraphics[width=0.475\textwidth]{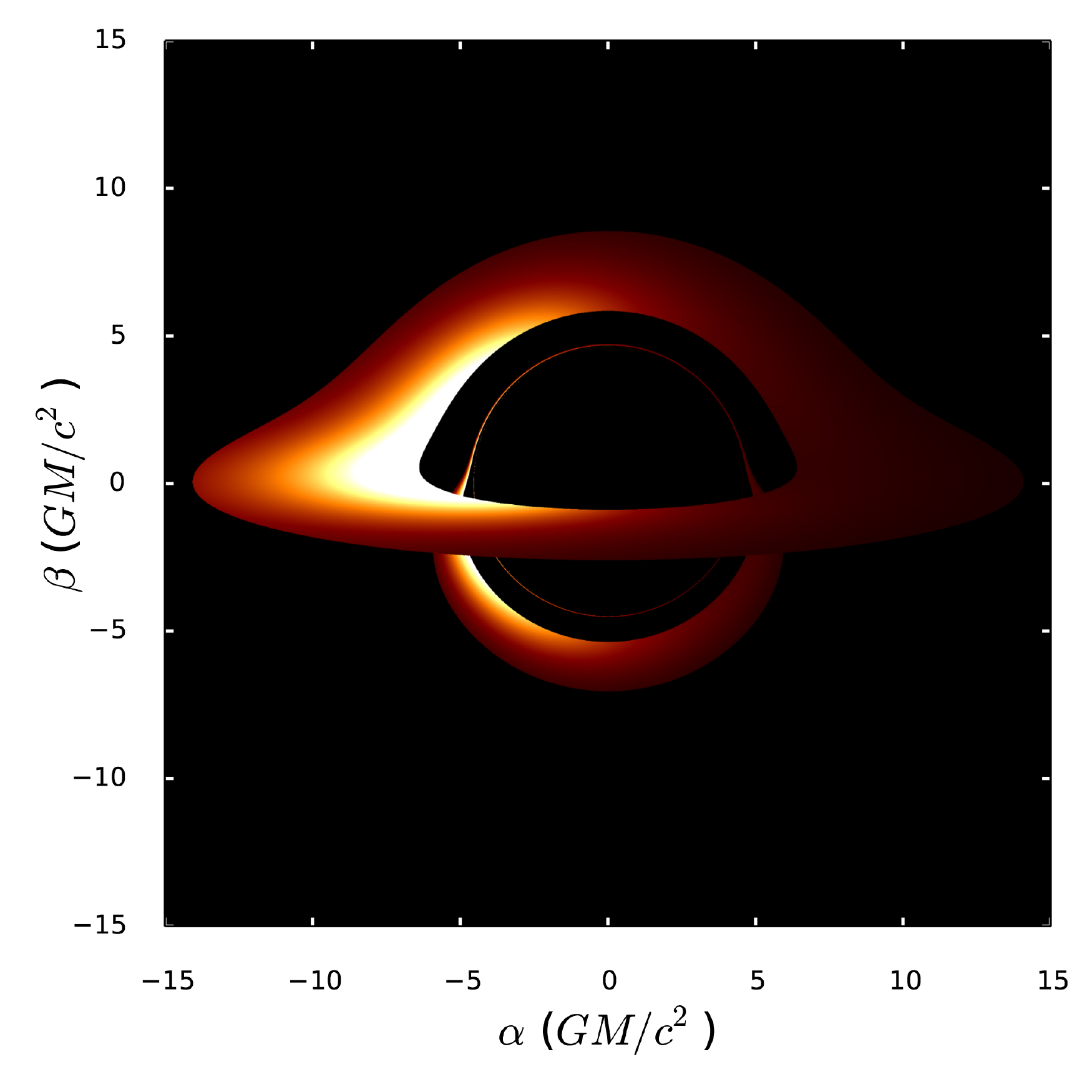}
\caption{Intensity map of the thin disk model described in Section
  \ref{sec:rad_trans}. Lensed images of up to tenth order are taken into
  account, where the order of an image is determined by the number of ray
  crossings of the equatorial plane. Higher-order images are ignored in
  the computation of the thin disk line spectra
  (Figs.~\ref{fig:thindiskspectra_a}--\ref{fig:thindiskspectra_d}). }
\label{fig:thindisk}
\end{figure}

The first model we investigated revolves around line emission profiles
emitted from a thin accretion disk, which has been studied by, among others,
\cite{luminet1979} and \cite{laor1991}, and reproduced
in~\cite{schnittman2004}, [see also \cite{Schnittman06},
  and~\cite{dexter2009}]. The model involves a steady state, optically
thick, geometrically thin accretion disk which is emitting monochromatic
radiation \citep{novikovthorne1973}. The disk extends from the black
hole's innermost stable circular orbit (ISCO), which depends on the black
hole spin $a$ \citep{bardeen1972}, to an outer radius of $15\,M$. The
emission coefficient is proportional to $1/r^2$ and there is no radiation
absorption. An image of the disk is shown in Fig.~\ref{fig:thindisk}
(note the effects of relativistic beaming, which brightens the parts of
the disk that move toward the observer). Due to relativistic beaming and
the gravitational redshift, the line emission is broadened. We computed the spectrum a distant observer would receive by recording the
intensity carried by each ray as well as its redshift (which is partly
gravitational and partly due to the disk's velocity).

\begin{figure*}
\centering
\begin{subfigure}[b]{0.475\textwidth}
	\includegraphics[width=\textwidth]{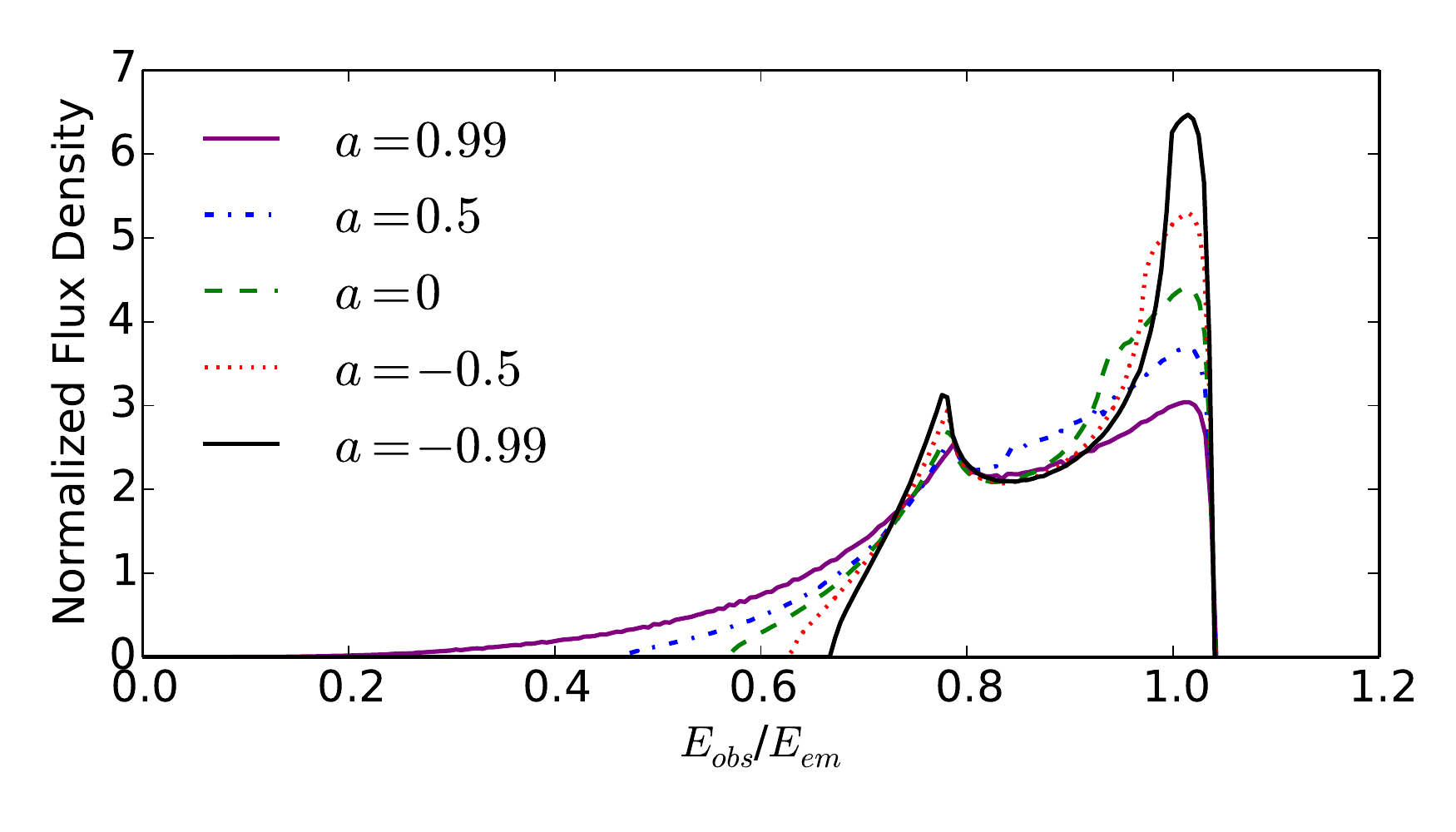}
	\caption{BL coordinates.}
    \label{fig:thindiskspectra_a}
\end{subfigure}
\begin{subfigure}[b]{0.475\textwidth}
	\includegraphics[width=\textwidth]{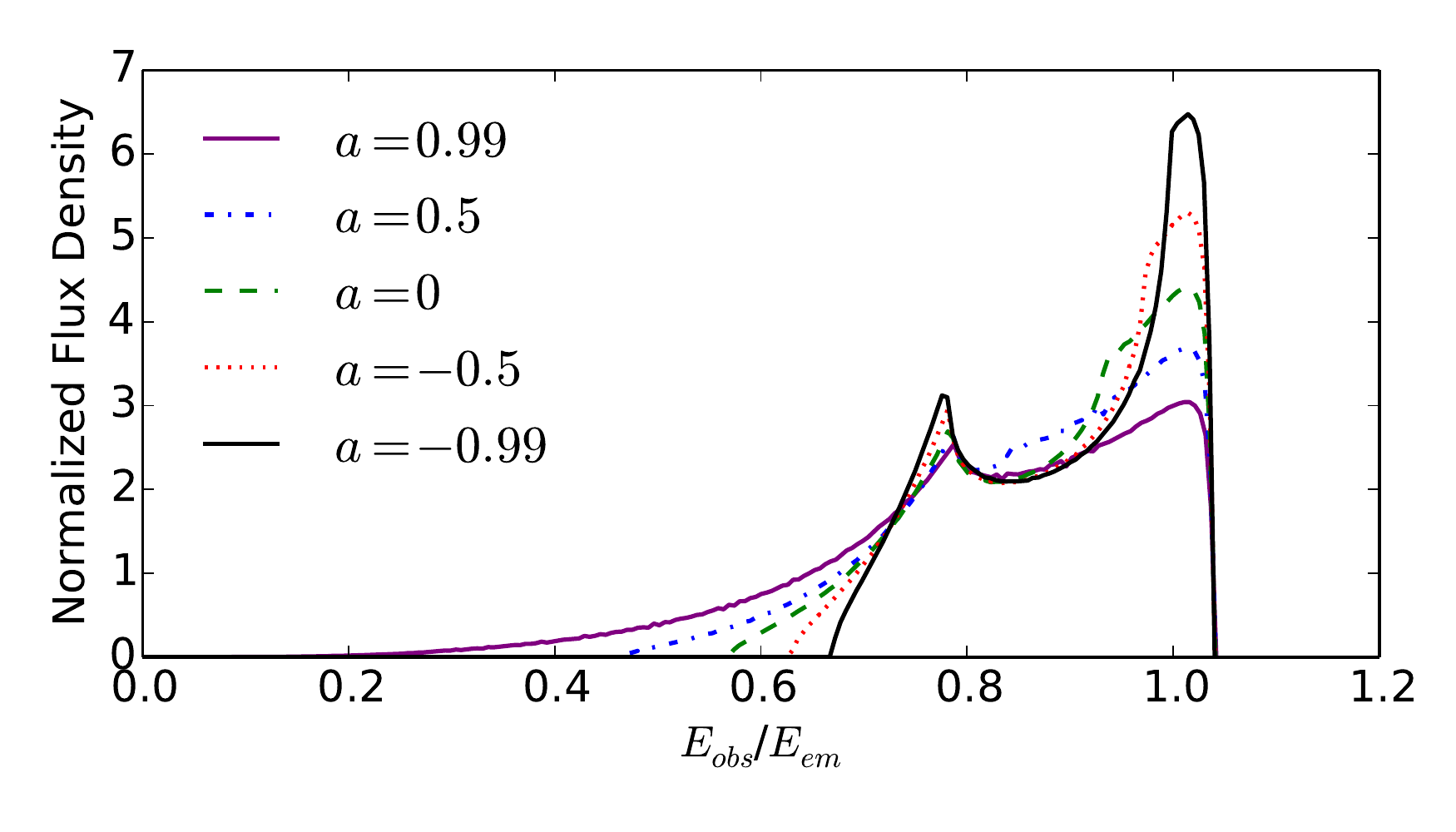}
	\caption{MBL coordinates.}
\end{subfigure}
\begin{subfigure}[b]{0.475\textwidth}
	\includegraphics[width=\textwidth]{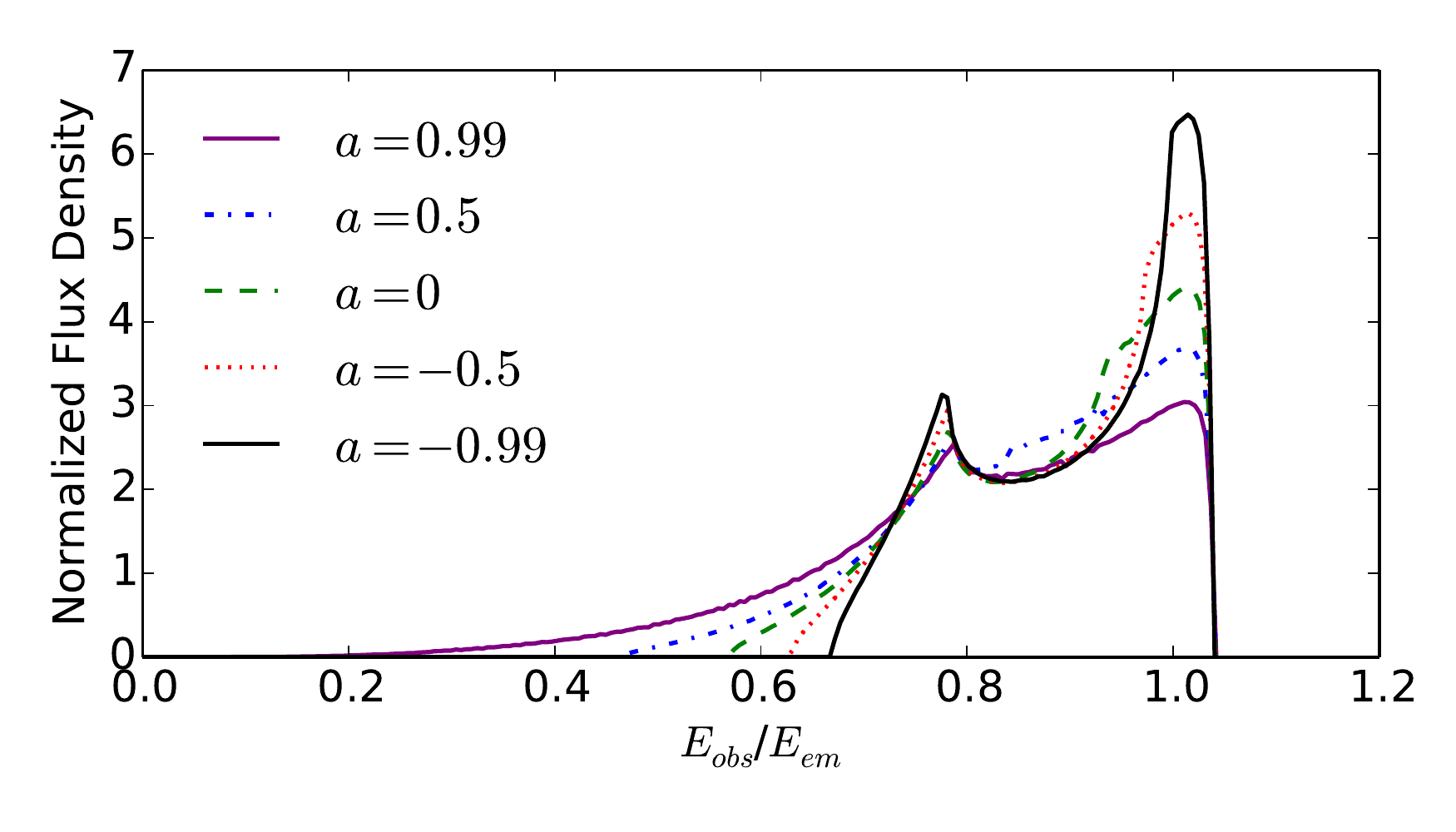}
	\caption{KS coordinates.}
\end{subfigure}
\begin{subfigure}[b]{0.475\textwidth}
	\includegraphics[width=\textwidth]{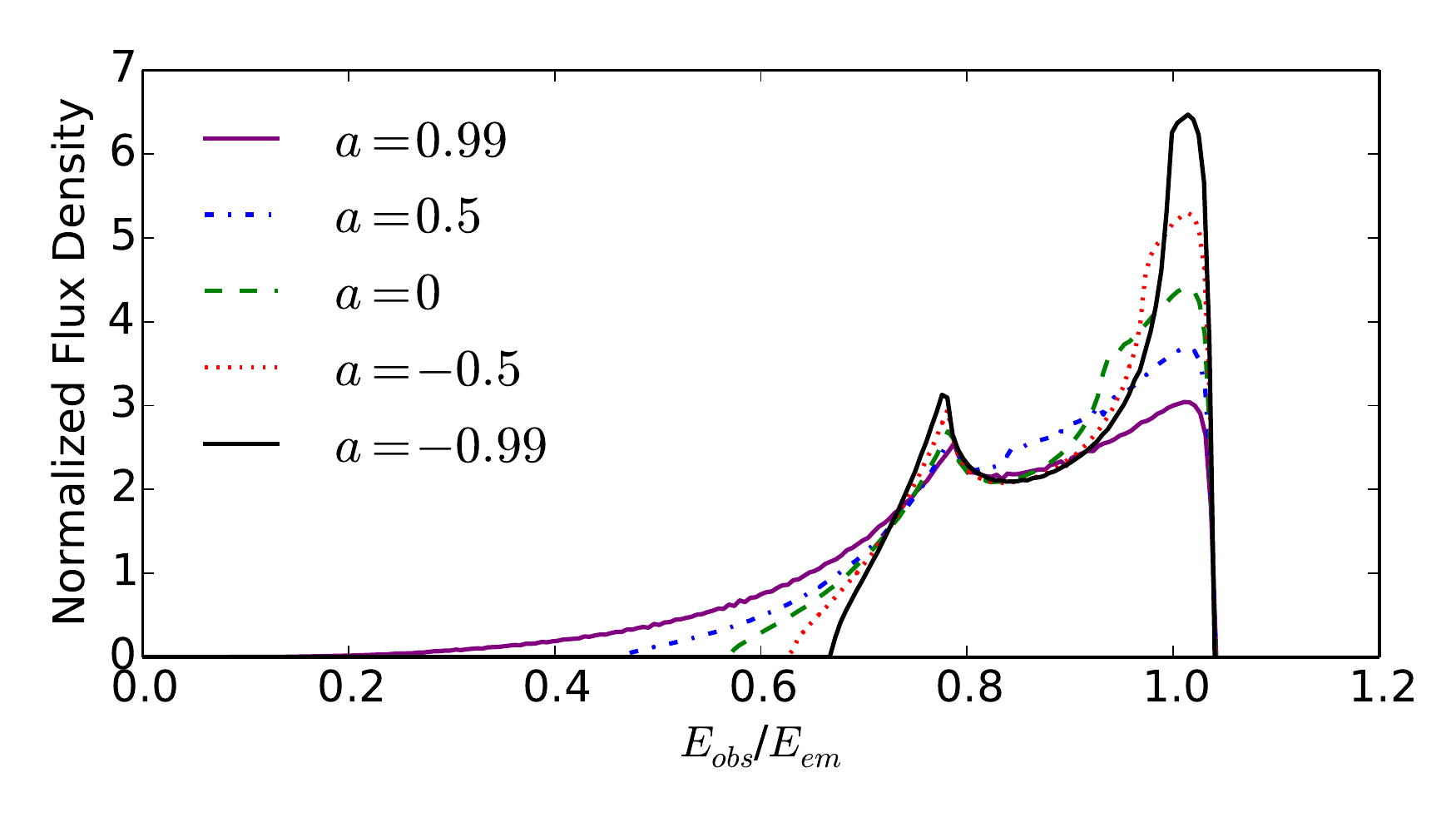}
	\caption{MKS coordinates.}
        \label{fig:thindiskspectra_d}
\end{subfigure}
\caption{Redshift spectra of the thin disk model described in Section
  \ref{sec:rad_trans} for four different coordinate systems. The five
  different spin parameters examined in these plots are (in descending
  order for the peaked curves above $E_{obs}/E_{em}=1$):
  $-0.99,-0.5,0,0.5,0.99$. The results may be compared with with
  ~\protect\cite{schnittman2004} and ~\protect\cite{dexter2009}.}
\end{figure*}

Figures \ref{fig:thindiskspectra_a}--\ref{fig:thindiskspectra_d} show the
spectra recorded by a distant observer for different black-hole spin
values (positive/negative spin values refer to prograde/retrograde disk
orbits, respectively). These results show a very good agreement with
Fig.~5 in \cite{dexter2009}.

\subsubsection{Synchrotron images of accretion flow simulated with {\tt
    BHAC}}
\label{sec:grmhd_convergence}

As a final test for our radiative-transfer integrator, we performed a
radiative-transfer simulation through GRMHD simulations of hot accretion
flow onto a SMBH with properties matching those observed in Sagittarius
A* (Sgr~A*). The simulations are produced using the {\tt BHAC}
three-dimensional (3D) GRMHD code \citep{BHAC}. During these calculations, the performance of {\tt RAPTOR} is monitored.

The dominant emission mechanism of many low-frequency (radio and
millimeter wavelengths) models of Sgr~A* is continuum synchrotron
emission from a relativistically hot magnetized plasma, that is, with
$\Theta_{\rm e} := k_BT_{\rm e}/m_{\rm e} c^2 \gtrsim 1$, with $k_B$,
$T_{\rm e}$, and $m_{\rm e}$ being the Boltzmann constant, the electron
temperature and mass, respectively. The synchrotron emission and
absorption coefficients depend on the electron distribution
function. Here, and in the remaining part the paper, we assumed that
the electrons have a thermal, relativistic (Maxwell-J{\"{u}}ttner)
distribution function:
\begin{equation}
n_{\rm e}^{TH}(\gamma)=\frac{n_{\rm e}^{TH} \gamma \sqrt{\gamma^2-1}
  \exp{\left(-\gamma/\Theta_{\rm e}\right)}}{ \Theta_{\rm e}
  K_2\left(1/\Theta_{\rm e}\right)}\,,
\end{equation}
where $\gamma$ is an electron's Lorentz factor and $K_2$ is the modified
Bessel function of the second kind. The normalization constant $n_{\rm
  e}^{TH}$ is the total electron number density, given by integrating the
distribution function over all possible electron Lorentz factors: $n_{\rm
  e}^{TH}=\int_1^{\infty} n_{\rm e}^{TH}\left(\gamma\right) d\gamma$.

The synchrotron emission coefficient $j_{\nu}$ of an ensemble of
electrons is obtained by integrating the synchrotron emission coefficient
of a single electron over the electron energies described by the above
distribution function. Since the above expression contains a Bessel
function, direct integration is time-consuming. In order to reduce the
required computing time for radiative-transfer calculations, we used an
approximate formula for the synchrotron emission coefficient provided by
\cite{leung2011}:
\begin{equation}
j^{TH}_\nu(\nu,\theta)=\frac{\sqrt{2} \pi {\rm e}^2 n_{\rm e}^{TH}
  \nu_s}{3 c K_2(\Theta_{\rm e}^{-1})} \left(X^{1/2}+2^{11/12}X^{1/6}\right)^2
\exp\left(-X^{1/3}\right)\,,
\label{eqn:synchemi}
\end{equation}
where $X$ is a dimensionless quantity given by
\begin{equation}
X:=\frac{\nu}{\nu_s},
\end{equation}
and $\nu_s$ is the critical frequency for the synchrotron emission:
\begin{equation}
\nu_s=\frac{2}{9} \left(\frac{eB}{2 \pi m_{\rm e} c}\right)
\Theta_{\rm e}^2 \sin \theta\,.
\end{equation}
Here, $B$ is the magnetic field in the inertial frame, $\theta$ is the
angle between the photon wave vector $k^\mu$ and the magnetic field
four-vector $b^\mu$ in the fluid frame, and $e$ is the unit electric
charge. The angle $\theta$ is calculated
by using Eq. 73 in \cite{dexter2016}. Here, we used CGS units, so that the
$B$ field is given in [Gauss] and $n_{\rm e}$ is given in $\rm [cm^{-3}]$. The
synchrotron emission coefficient therefore has units of ${\rm [ergs \,
    s^{-1} Hz^{-1} cm^{-3}]}$. In this test, we assumed that
$K_2\left(\Theta_{\rm e}^{-1}\right)=2\,\Theta_{\rm e}^2$, which is an
approximation. However, evaluating properly the Bessel function can yield
differences in the integrated flux on the order of a few percent at most.

As shown in \cite{leung2011}, Eq.~\eqref{eqn:synchemi} is a good
approximation of the true synchrotron emission coefficient for a range of
electron temperatures $\Theta_{\rm e}$ and frequencies $\nu$. For
$\theta=30\degr$, the relative error of the approximate emission
coefficient formula is less than 1\% for $\Theta_{\rm e} > 0.5$, and
$\nu/\nu_c>10$ (where $\nu_c=eB/2 \pi m_{\rm e} c=2.8 \times 10^6 B$
Hz is the electron cyclotron frequency). For $\theta=80\degr$, the error
is less than 1\% when $\Theta_{\rm e} >0.5$ and $\nu/\nu_{\rm
  c}>10^4$. In our models, the typical magnetic field strength is
$B\simeq 10$ Gauss, so $\nu_{\rm c}\simeq 10^7$ Hz. Hence, when modeling
the emission at frequencies around $\nu=10^{11}$ Hz, the approximate
synchrotron emission coefficient formula given by
Eq.~\eqref{eqn:synchemi} is very accurate.

The synchrotron self-absorption coefficient for a thermal distribution of
electrons is derived from Kirchhoff's law, $\alpha^{TH}_\nu =
j^{TH}_{\nu}/B_{\nu}$, where $\alpha^{TH}_\nu$ is given in $\rm
[cm^{-1}]$ and ${B}_\nu$ is the Planck function:
\begin{equation}
{B}_{\nu}:=\frac{ 2 h \nu^3}{c^2} \frac{1}{\exp{\left( h\nu/ (m_{\rm e}
    c^2 \Theta_{\rm e}) \right)} - 1}\,.
\end{equation}

{\tt BHAC} (and similar GRMHD codes such as {\tt HARM}) employ
geometrized units, so that to carry out radiative-transfer calculations
using the GRMHD simulation data for the plasma variables we must scale
the GRMHD variables using the rest-mass density scaling factor
$\rho_0:=\mathcal{M}/\mathcal{L}^3$ and the magnetic field strength
scaling factor $B_0:=c\sqrt{4 \pi \rho_0}$. Here, ${\mathcal L}=\rg$ and
${\mathcal T} = \tunit$ are the simulation's length and
time scale factors, respectively; they are functions of the black hole's mass only. The mass unit, ${\pazocal M}$, is a free parameter of
the model. The electron number density used in the synchrotron
emission/absorption coefficient formulae is then calculated using
\begin{equation}
 n_{\mathrm{e}} = \rho \frac{\rho_0}{(m_{\rm e}+m_{\rm p})} \,
 \mathrm{cm}^{-3}\,.
 \end{equation}
In our GRMHD model, the temperature of protons $T_{\rm p}$ is
proportional to the ratio of the plasma's pressure and density, so that
it is a scale-free quantity. The electron temperature $T_{\rm e}$, which
is essential to calculating the synchrotron emissivities, is parametrized
by the proton-to-electron temperature ratio, which can be either constant
or a function of GRMHD data variables. The key parameters for the test
simulation are listed in Table \ref{table:convtest}, where
$\tau_{\mathrm{cutoff}}$ represents the optical depth at which
integration is terminated (radiation emitted past this optical depth will
have a negligible effect on the image).

\begin{table}
\centering
\begin{tabular}{l l}
\hline
Variable & Value \\
\hline
Mass & $4.5 \times 10^6\,M_{\odot}$ \\
Distance 	   & $8.5\,{\rm kpc}$ \\
$r_{\mathrm{camera}}$ & $10^4\,M$  \\
Range for $\alpha$ & $[-30, 30]\,M$ \\
Range for $\beta$  & $[-30, 30]\,M$ \\
Resolution $(x)$   & $16,32,64,128,256,512,1024,2048,4096$ px \\
Resolution $(y)$   & $16,32,64,128,256,512,1024,2048,4096$ px \\
Frequency 	   & $230\,{\rm GHz}$ \\
Inclination 	   & $90\degr$ \\
$\tau_{\mathrm{cutoff}}$ & $\ln(1000)$ \\
$T_{\rm p}/T_{\rm e}$ 	   & $3.0$ \\
\hline
\end{tabular}
\caption{Setup for the comparison test between {\tt BHOSS} and {\tt RAPTOR}. The mass and distance estimates for Sgr A* used for our convergence test were obtained from \citep{ghez2008}.}\label{table:convtest}
\end{table}

It is found that, when using 10 CPU cores, {\tt RAPTOR} integrates 10,707
geodesics per second, while with one GPU unit, {\tt RAPTOR} integrates
104,900 geodesics per second. The total run time for an image of 2000 $\times$ 2000 pixels was less than one minute using a GPU. A more detailed description of the code performance is given in Appendix \ref{sec:code_performance}.

Figure~\ref{fig:BHOSSRAPTOR} (top left panel) presents an image computed by
{\tt RAPTOR} of our GRMHD accretion flow simulation created in {\tt
  BHAC}. We compared our results to images produced by another radiative
transfer code, {\tt BHOSS} \citep{BHOSS} (top right panel), which uses a different set of
algorithms to {\tt RAPTOR}. The total fluxes from both
codes as a function of image resolution are given in
Table~\ref{table:convergence}. {\tt BHOSS} and {\tt RAPTOR} converge to
almost identical total flux values, with a relative difference of
$0.06\%$ at a resolution of $4096 \times 4096$ pixels. 
The percentage difference for every pixel in both images is presented in the bottom left panel, while the absolute difference between both images is shown in the bottom right panel. The overall structure of the images is consistent; the largest percentage differences are found in regions of low specific intensity, whose contribution to the observed flux density is negligible (see bottom right panel).

A possible source of the differences between {\tt BHOSS} and {\tt RAPTOR}
is the fact that the two codes employ different approaches for computing
step-sizes. {\tt BHOSS} uses an algorithm that depends on the optical
depth of the plasma, while {\tt RAPTOR} bases its step-size purely on the
geometry of spacetime. This results in different sampling strategies
which, although not a dominant factor in the total flux, can nonetheless
result in large deviations in a plot of the relative difference between
the two images. Looking at the images and total fluxes, we conclude that the codes give
consistent results while using different methods to solve the radiative
transport and geodesic equations. The GRMHD file used in this comparison
is included with the {\tt RAPTOR} code.

\begin{figure*}
\centering
\includegraphics[width=0.8\textwidth]{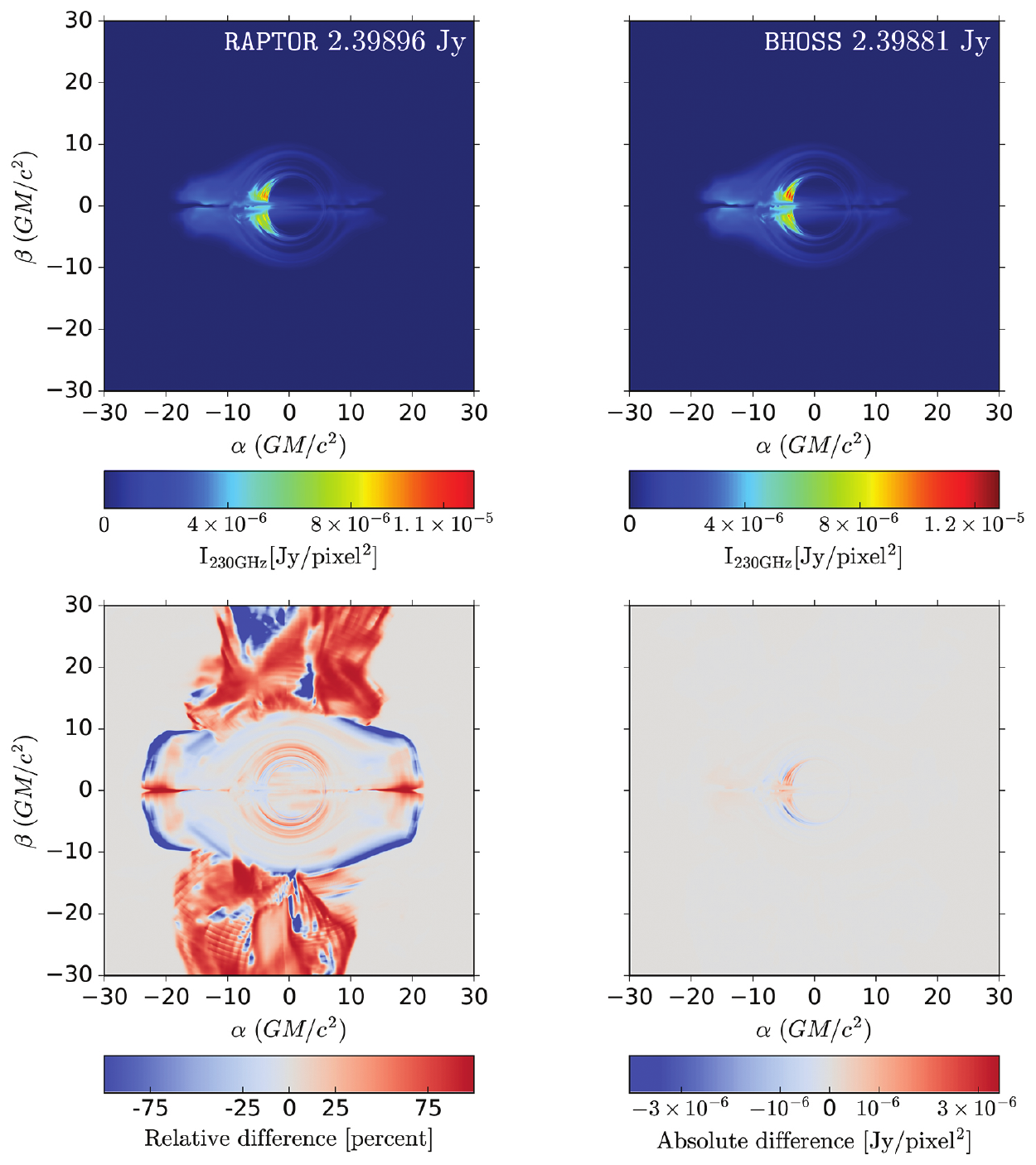}
\caption{Intensity maps at $230$ GHz for the comparison test, with a
  resolution of $4096 \times 4096$ pixels. The {\tt RAPTOR} output 
  (top left panel) and {\tt BHOSS} output (top right panel) show
  excellent agreement in total flux density. The relative difference between the output of both codes is plotted
  in the bottom left panel. The deviations become quite large in the periphery of the image; since the intensity in these regions is low, however, they do not contribute significantly to the integrated flux density, as can be seen in the absolute difference between the two codes (bottom right panel). }\label{fig:BHOSSRAPTOR}
\end{figure*}

\begin{table}
\centering
\begin{tabular}{c c c c c c}
\hline
\hline
pixels &	$I_{\rm 230\ GHz,\, RAPTOR}$	& $I_{\rm 230\ GHz,\,
  BHOSS}$	& 	$\Delta I/I$ \\
\hline
$128 \times 128$	&	2.39467	&		2.36903	&		1.07  \\
$256\times 256$   &	2.39794	&    	2.38237	&		0.65 \\
$512\times 512$   &	2.39871	&		2.39133	&		0.31 \\
$1024 \times 1024$	&	2.39899	&		2.39553	&		0.14 \\
$2048 \times 2048$	&	2.39898	&		2.39777	&		0.050\\
$4096 \times 4096$	&	2.39896	&		2.39881	&		0.0063 \\
\hline
\end{tabular}
\caption{Integrated flux density computed by {\tt RAPTOR} and {\tt BHOSS}
  for the model described in Section \ref{sec:grmhd_convergence}, as well
  as the relative error, showing convergence of the output of the two
  codes. The last column reports the relative difference between the two
  codes, which is, $ \Delta I/I := ({I_{\rm 230\ GHz,\, RAPTOR} - I_{\rm
      230\ GHz,\, BHOSS}})/{I_{\rm 230\ GHz,\,
      RAPTOR}}$.}\label{table:convergence}
\end{table}

\section{Time-dependent radiative transfer in GRMHD simulations}
\label{sec:slowlight}

GRMHD simulations dump their data in so-called snapshots - instantaneous states
of the plasma captured at discrete moments during the simulation. When
creating images based on a series of GRMHD snapshots using
radiative-transfer calculations, it is either possible to ignore a
geodesic's time coordinate, thereby treating the GRMHD snapshot as static
while the rays propagate (this is normally referred to as the
fast-light approximation), or it is possible to keep track of
the geodesic's time coordinate, interpolating between different GRMHD
snapshots so as to obtain the plasma conditions that are correct not only
in space but also in time (this is normally referred to as the
slow-light approximation).

In order to generate the results of this section, we relaxed the assumption of staticity of the GRMHD
simulation during radiative-transfer calculations and thus implement the
slow-light approximation with the goal of improving the accuracy of our
model in predicting the properties of the accretion flow onto a SMBH. In
such relativistic plasmas, the effects of selecting the fast-/slow-light
approximation can be arbitrarily great or small based on the geometry and
observer under consideration. For instance, \citet{dexter2010} considered
the differences between fast-light and slow-light in the context of a
particular accreting black-hole model and concluded that the differences
were minimal. Here, we return to this problem as new models with more
complex electron thermodynamics have been developed. Overall, we are here
interested in verifying the fast-light approximation in a broader
context.

In GRMHD simulations such as those performed by {\tt HARM2D} or {\tt
  BHAC}, only the heavy ions are simulated. In radiative processes,
however, ions are not the dominant source of emission, and we therefore
need a description to couple the electrons in the plasma to their ionic
counterparts. To do so, we implemented a one-fluid model in which the
coupling between the two species changes throughout the simulation volume
as a function of the ${\beta_{\rm p}}$ plasma parameter \citep{moscibrodzka2016}
\begin{equation}
\frac{T_{\mathrm{p}}}{T_{\mathrm{e}}} = 
R_{\rm low}\frac{1}{1+\beta_{\rm p}^2} +
R_{\rm high} \frac{\beta_{\rm p}^2}{1+\beta_{\rm p}^2}\,,
\end{equation}
where $\beta_{\rm p} := {P_{\rm gas}}/{P_{\rm mag}}$ is the ratio of gas pressure
to magnetic-field pressure $P_{\rm mag}=B^2/2$, and $R_{\rm low}$ and
$R_{\rm high}$ are two free parameters. In strongly magnetized plasmas,
$\beta_{\rm p} \ll 1$ and $\trat \rightarrow R_{\rm low}$. In weakly magnetized
plasmas, on the other hand, $\beta_{\rm p} \gg 1$ and so $\trat \rightarrow
R_{\rm high}$. 

The models used in this section are motivated by qualitatively
reproducing the observed spectrum of Sgr A* \citep{shcherbakov2012}. Our
primary focus, however, is on investigating the difference (if any)
between the fast-light and slow-light paradigms, rather than on
reproducing the source in great detail. More precisely, the parameters
assumed for our modeling are that the SMBH has a mass of $M=4 \times 10^6
M_{\odot}$ and is at a distance of $d=7.88 \times 10^{3}\,{\rm pc}$
\citep{boehle2016}, that the observer has an inclination of $60\degr$
and performs observations at frequencies of $22, 43, 86$, and $230\,{\rm
  GHz}$.

\subsection{Time calibration}

When comparing slow and fast-light simulations it is important to
describe how the results align when comparing them. The reason for this
is that a single image from a slow-light calculation uses data from many
GRMHD slices, so that it is no longer possible to associate an image with
a specific GRMHD slice, as is possible in the case of a fast-light
calculation. This also means that a large number of GRMHD slices are
needed for slow-light calculations: the slices of GRMHD data should
extend from the moment in time when the first ray enters the simulation
volume until the moment when the last ray leaves it. When taking
curvature into account, these moments are difficult to compute and depend
on the initial conditions of the camera rays: some images can contain
rays that pass close to the event horizon, circling the black hole many
times before escaping (see Fig.~\ref{fig:diff_geo}), thus delaying their
exit time, potentially indefinitely for pathological rays. We here
neglected the contribution of these rays for two reasons; first, the
optical depth along such a ray would cause virtually all emission beyond
a certain value for the affine parameter to be absorbed; second, rays
that orbit the black hole for a an increasing number of times are
increasingly rare, thus contributing only marginally to our images.

In practice, we established the alignment by introducing an empirically
determined time delay in our rays' initial time coordinate in such a way
that light curves from the fast-light and slow-light simulations at an
inclination angle of $90\degr$ coincide; this is effectively equivalent
to translating the slow-light light curve with respect to the fast-light
light curve on the time axis.

\subsection{Results}

\begin{figure*}
\begin{center}
\includegraphics[width=0.40\textwidth]{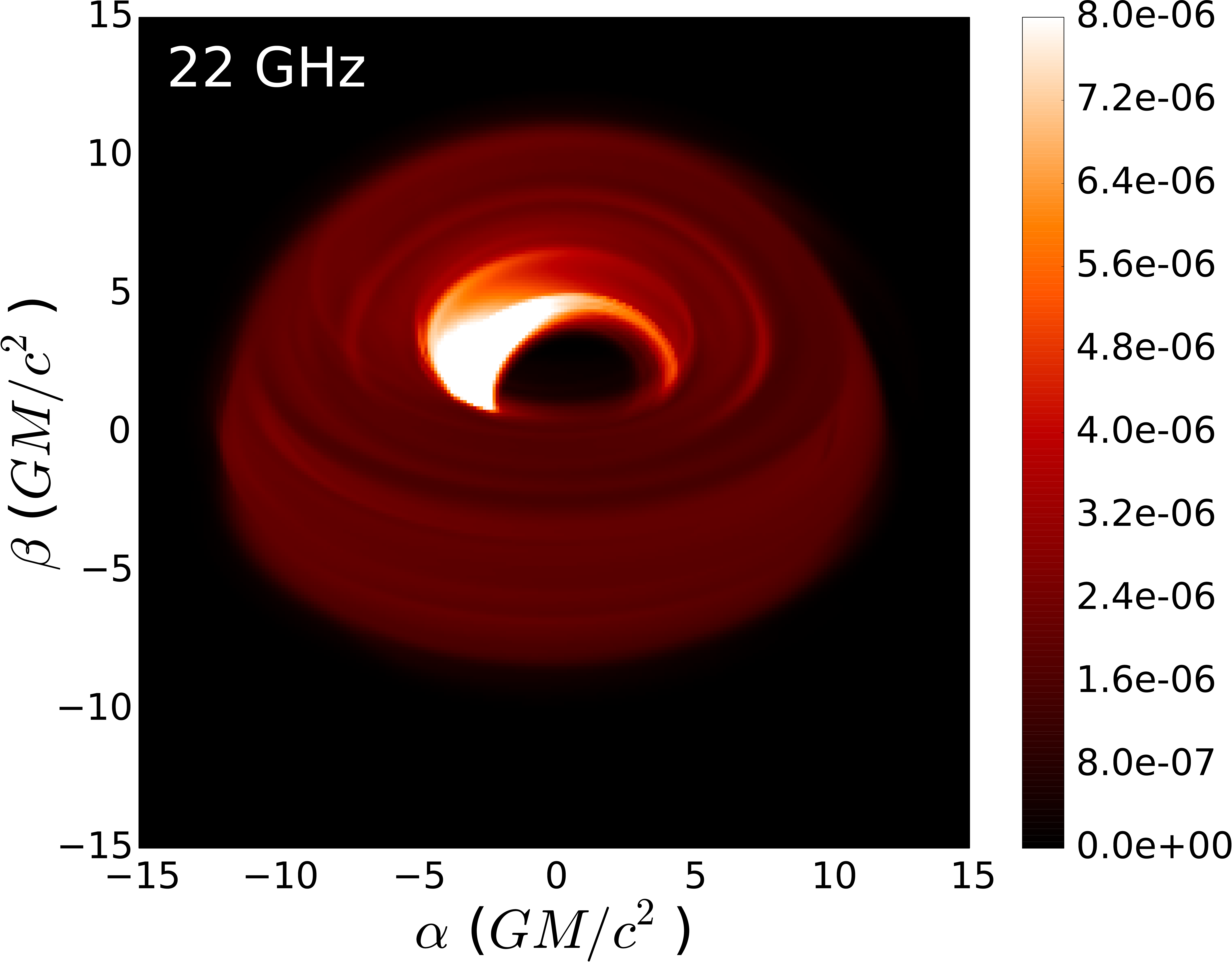}
\includegraphics[width=0.40\textwidth]{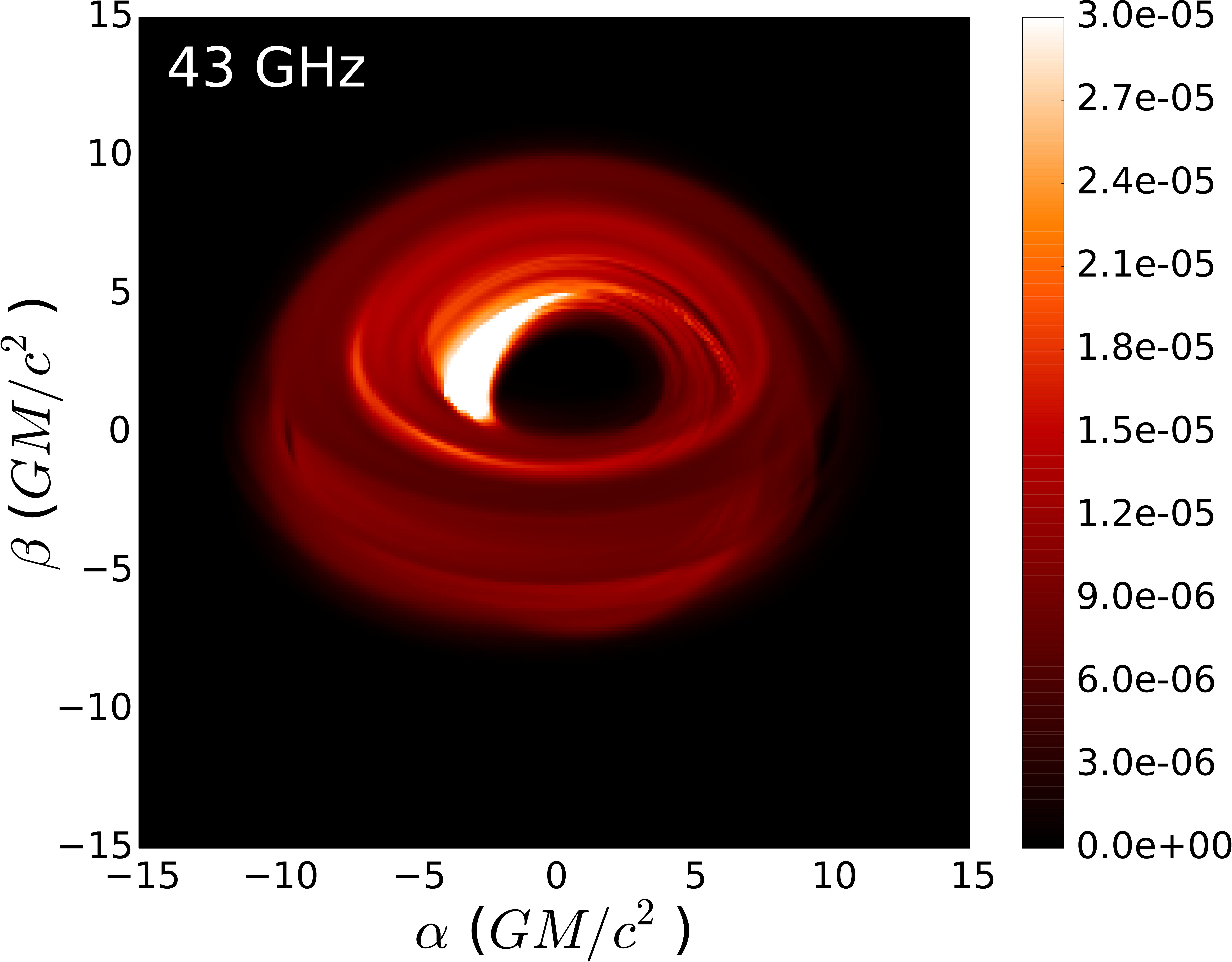}\\
\vskip 0.5cm
\includegraphics[width=0.40\textwidth]{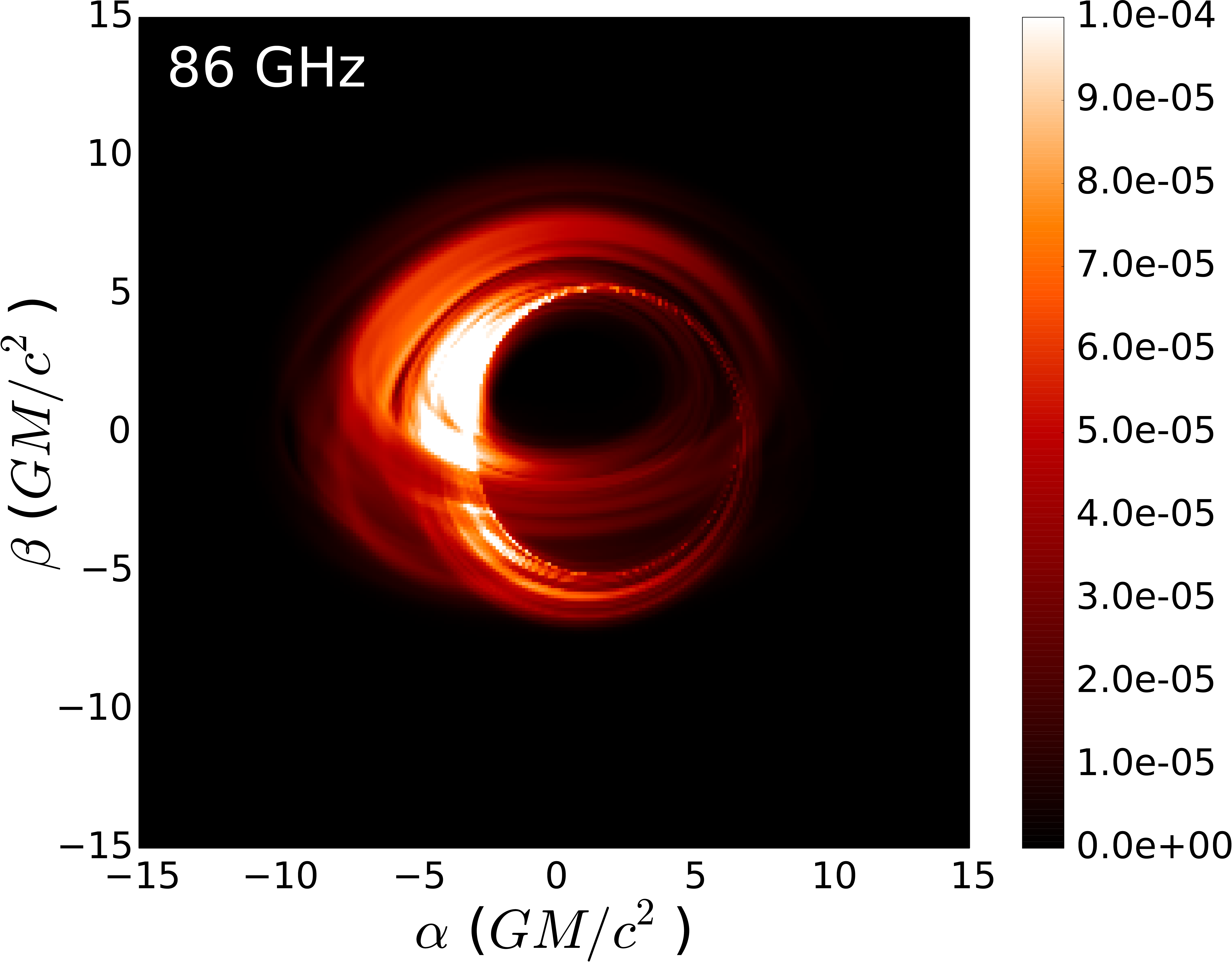}
\includegraphics[width=0.40\textwidth]{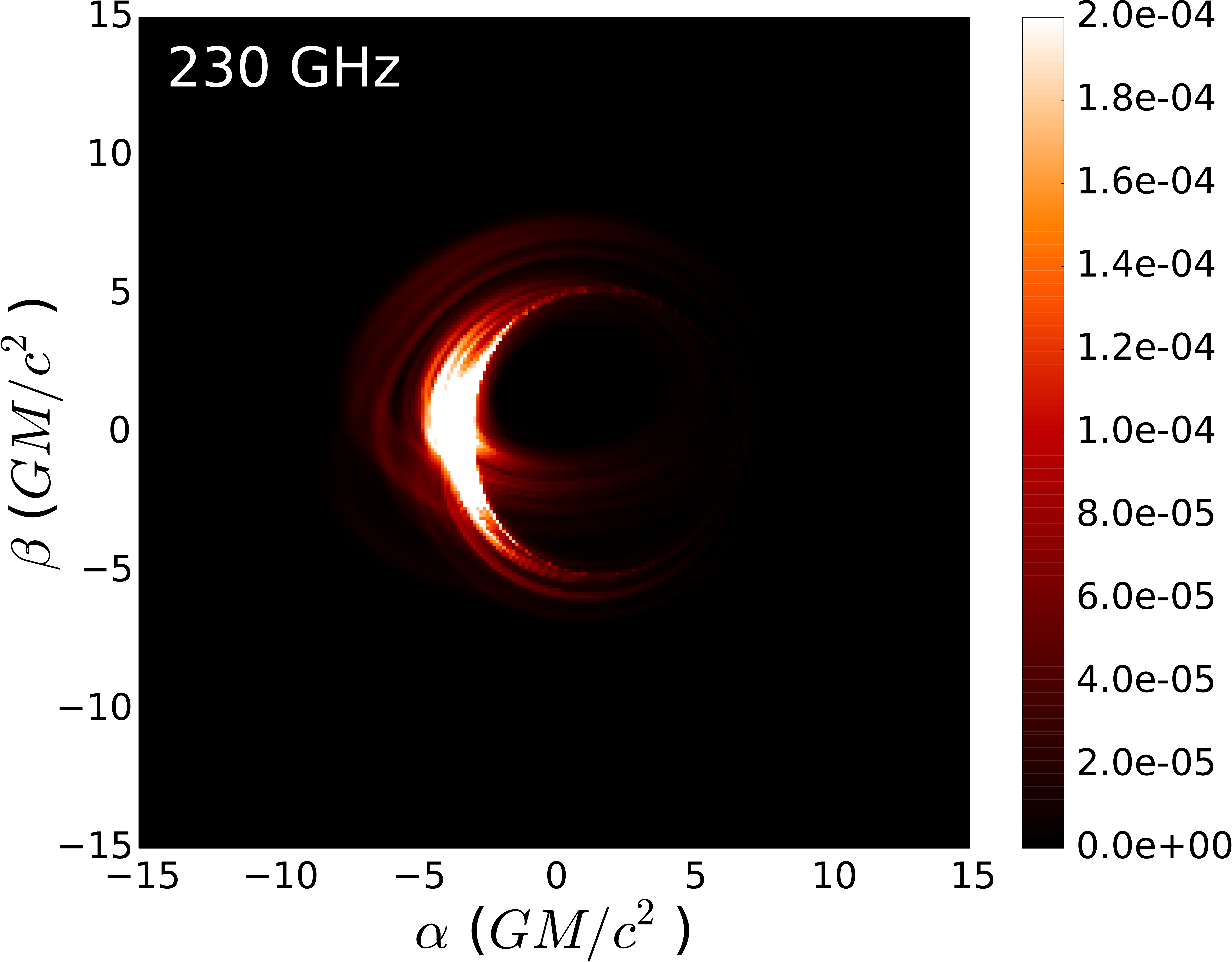}
\end{center}
\caption{Images of our slow-light-simulation of the disk emission
  dominated model at VLBI frequencies. The flux density is given in units
  of Jy pixel$^{-2}$. }
\label{fig:disk230}
\end{figure*}

\begin{figure*}
\begin{center}
\includegraphics[width=0.45\textwidth]{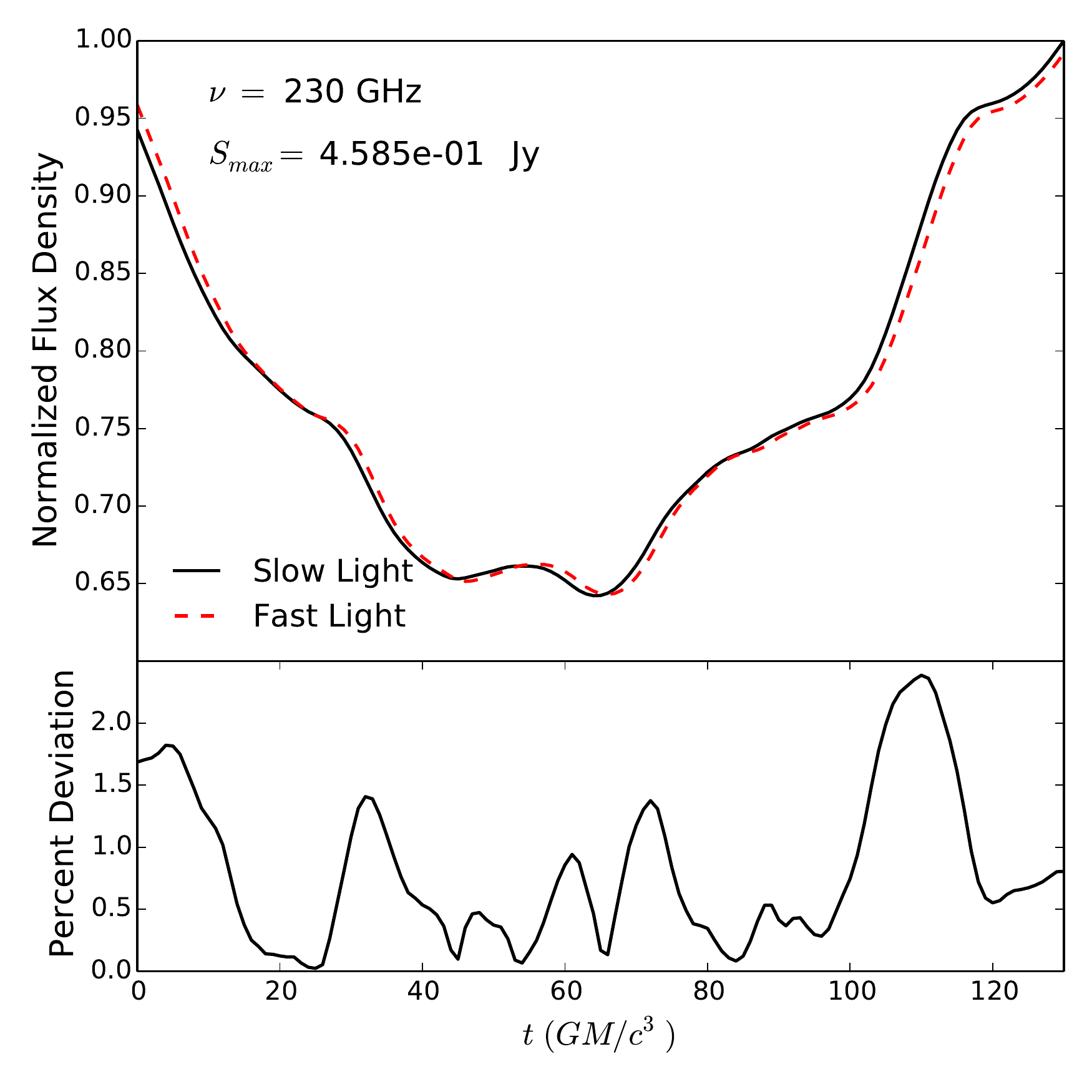}
\includegraphics[width=0.45\textwidth]{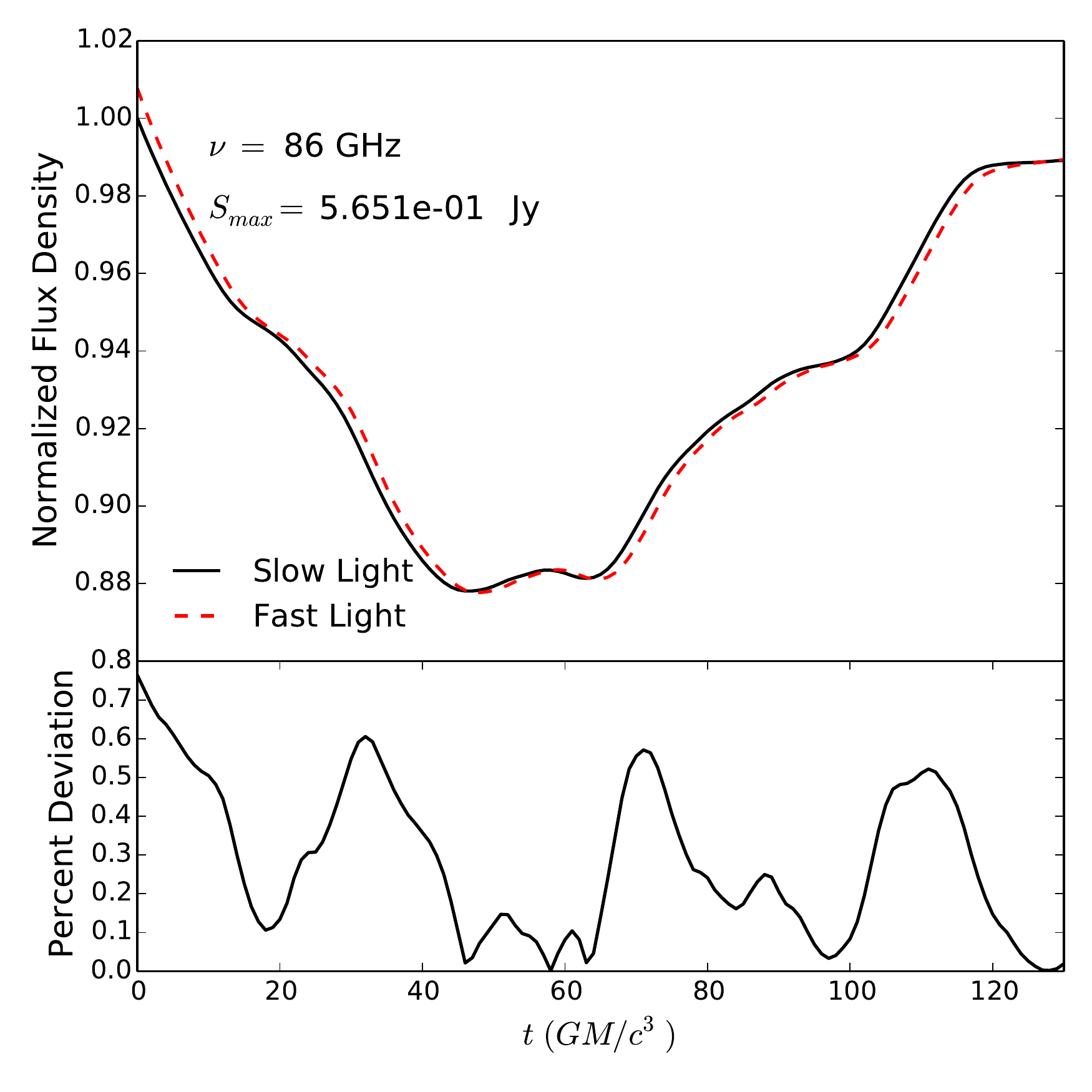}\\
\includegraphics[width=0.45\textwidth]{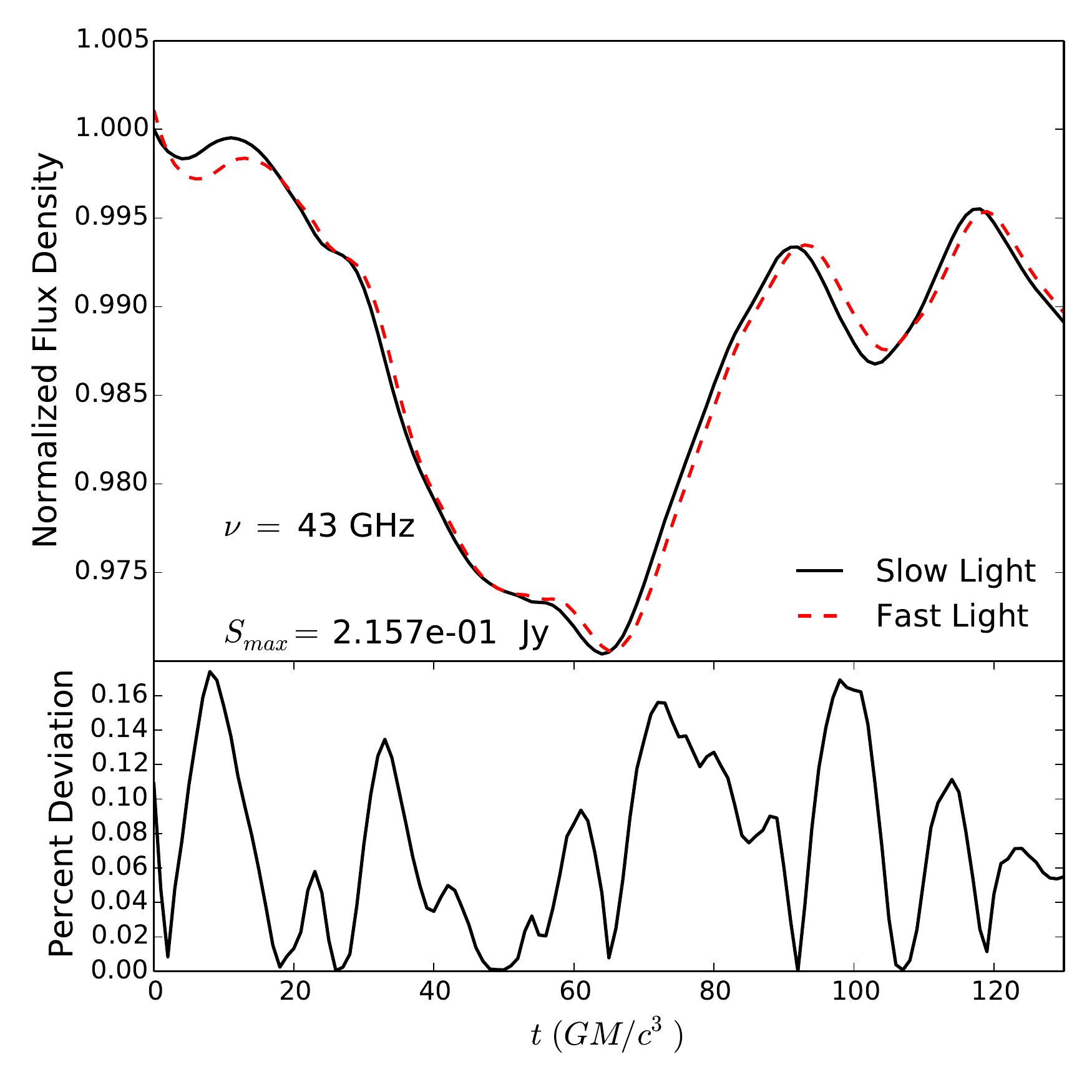}
\includegraphics[width=0.45\textwidth]{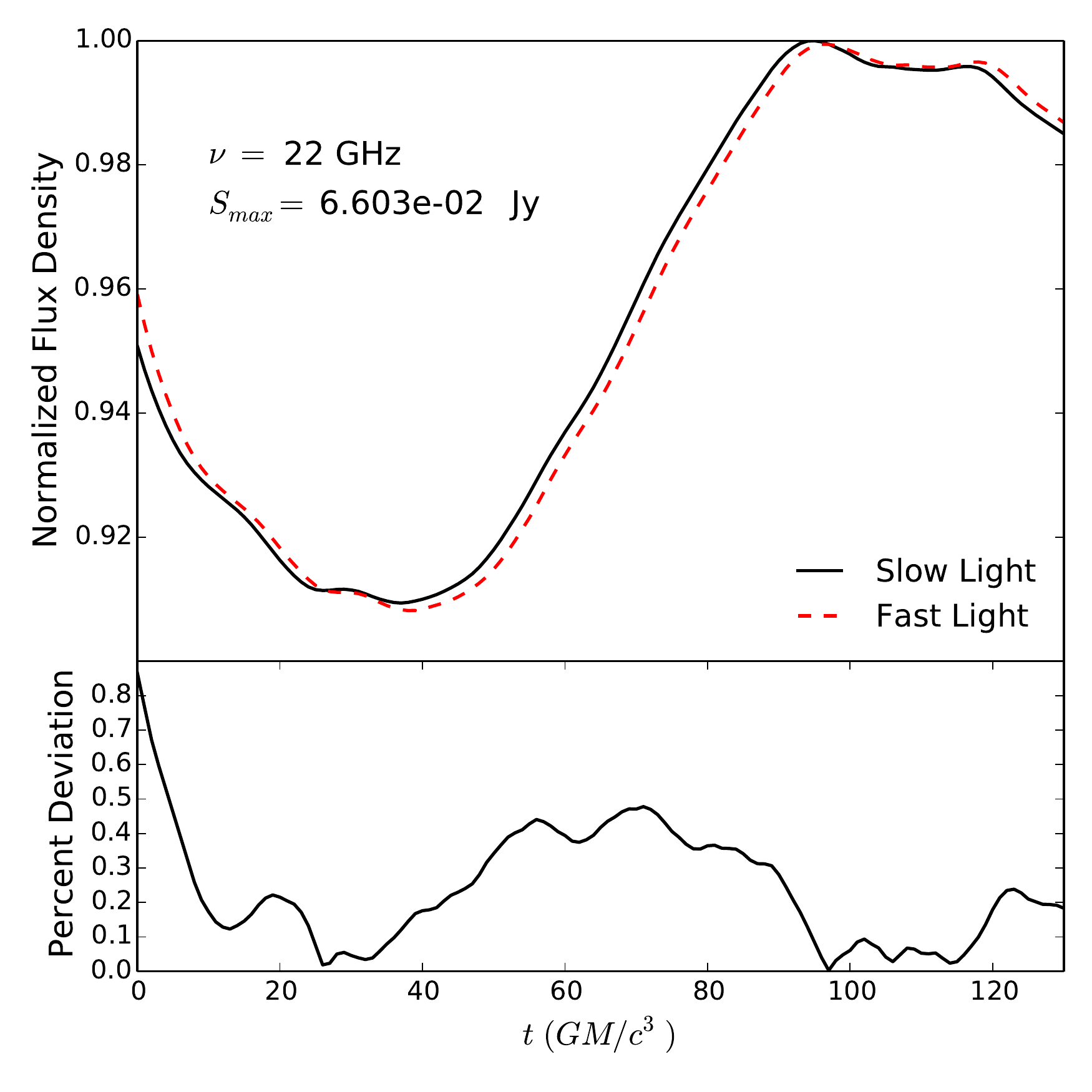}
\end{center}
\caption{Light curves for slow-light (solid) and fast-light (dashed) for
  the disk emission dominated model at VLBI frequencies. Residuals are
  displayed below the light curves.}
	\label{fig:lightcurves_disk_230GHz_dashed_FL}
\end{figure*}

\begin{figure*}
\begin{center}
\includegraphics[width=0.70\textwidth]{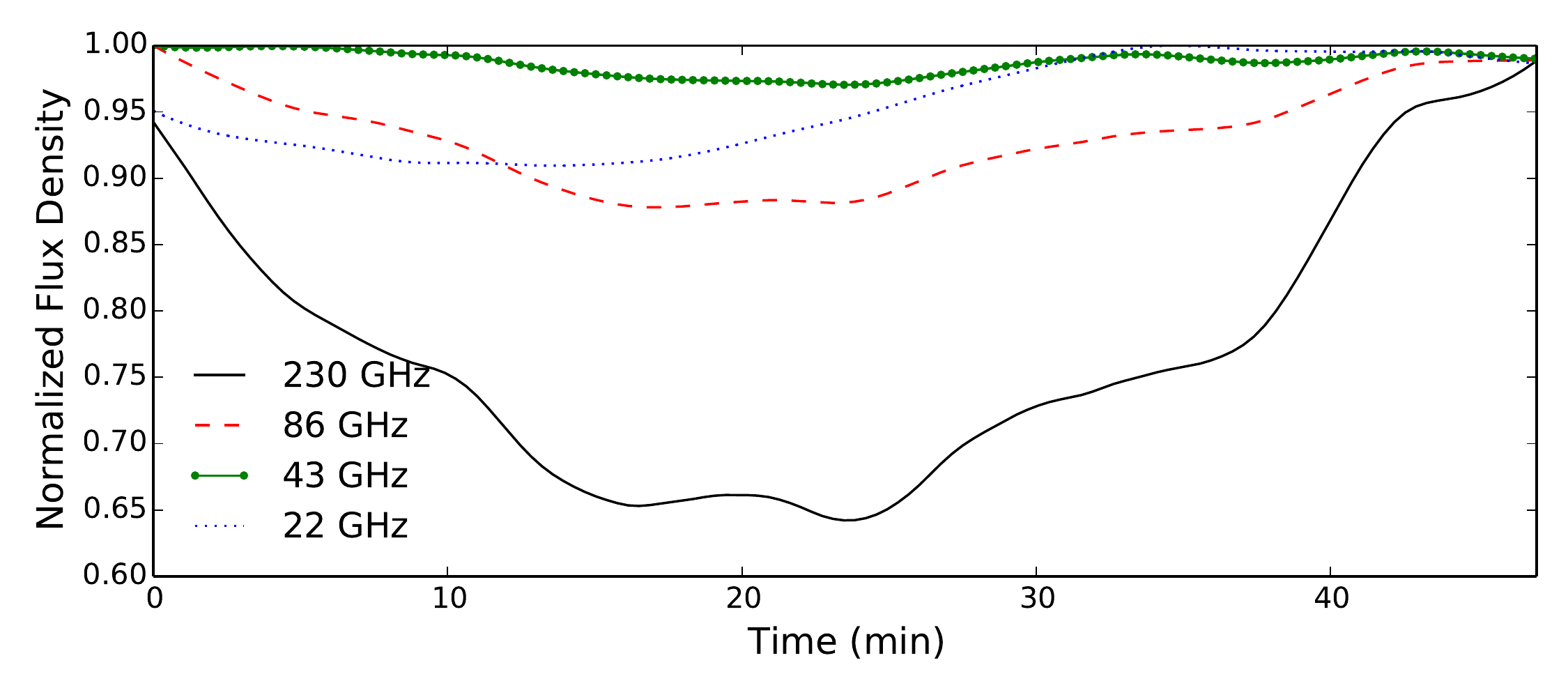}
\caption{Normalized light curves of slow-light simulations at the VLBI
  frequencies (230 GHz (solid), 86 GHz (dashed), 43 GHz (solid and
  dotted), 22 GHz (dotted)) for viewing angle of 60 degrees for our disk
  dominated model.}~\label{fig:multifreq60deg_disk}
\end{center}
\end{figure*}

The model whose emission is dominated by the disk is characterized by
a mass scale factor, $\mathcal{M}=5.03 \times
10^{-15}\,M_{\odot}$ and the weak-magnetization
electron-proton temperature ratio $R_{\rm high}=R_{\rm low}=1$. Slow-light images of
the accretion disk at our 4 VLBI frequencies are presented in
Fig.~\ref{fig:disk230} (we do not report the corresponding fast-light
images as they are virtually indistinguishable). Light curves for both
the slow-light and fast-light simulations, along with residuals, are
presented in Fig.~\ref{fig:lightcurves_disk_230GHz_dashed_FL}, while
normalized light curves at all frequencies for the slow-light simulation
of the disk-emission model are shown in
Fig.~\ref{fig:multifreq60deg_disk}.

\begin{figure*}
\begin{center}
\includegraphics[width=0.40\textwidth]{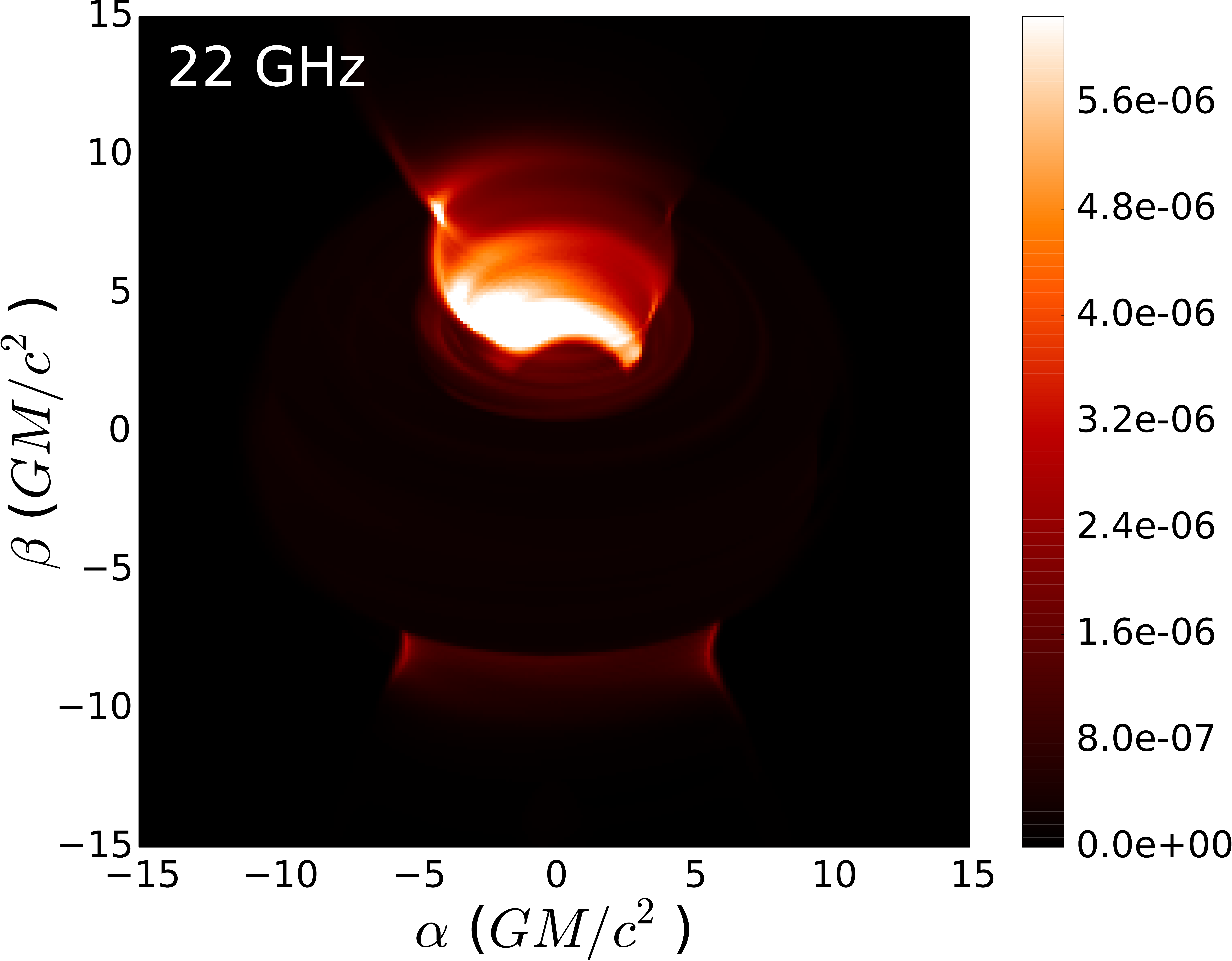}
\includegraphics[width=0.40\textwidth]{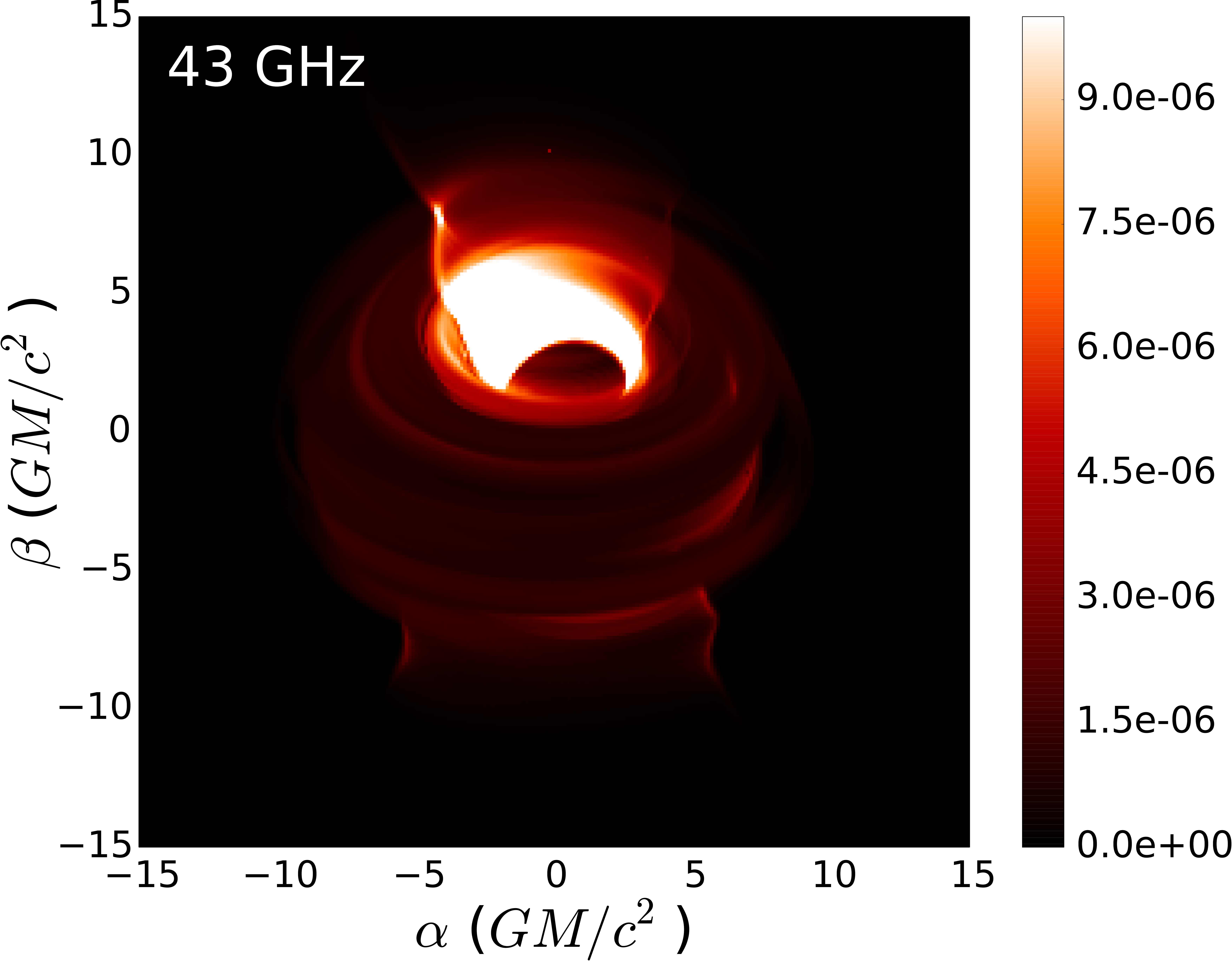}\\
\includegraphics[width=0.40\textwidth]{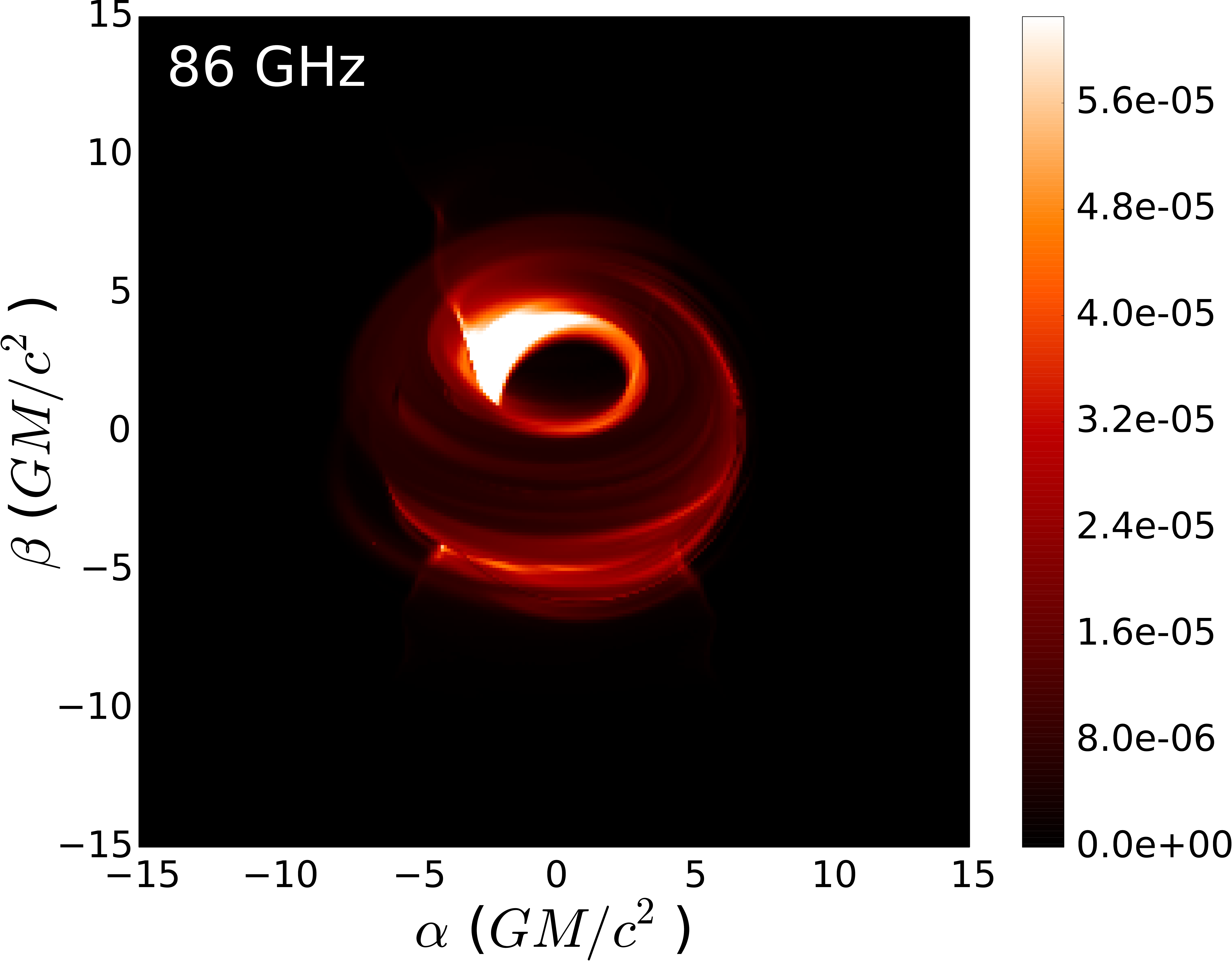}
\includegraphics[width=0.40\textwidth]{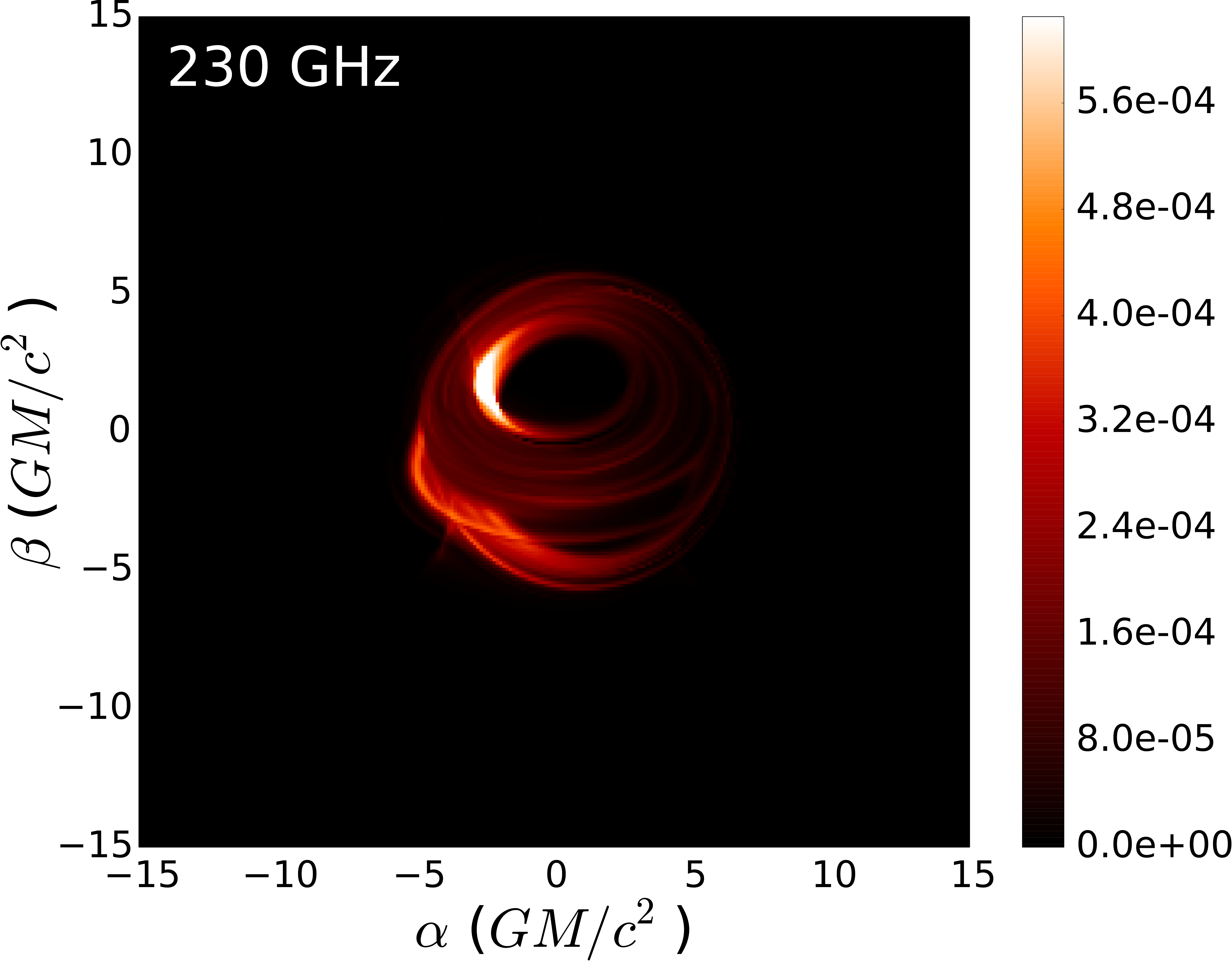}
\end{center}
\caption{Images of our slow-light-simulation of the jet emission
  dominated model at VLBI frequencies. Flux density is given in units of
  Jy pixel$^{-2}$.}~\label{fig:230GHz_jet_SL}
\end{figure*}

\begin{figure*}
\begin{center}
\includegraphics[width=0.45\textwidth]{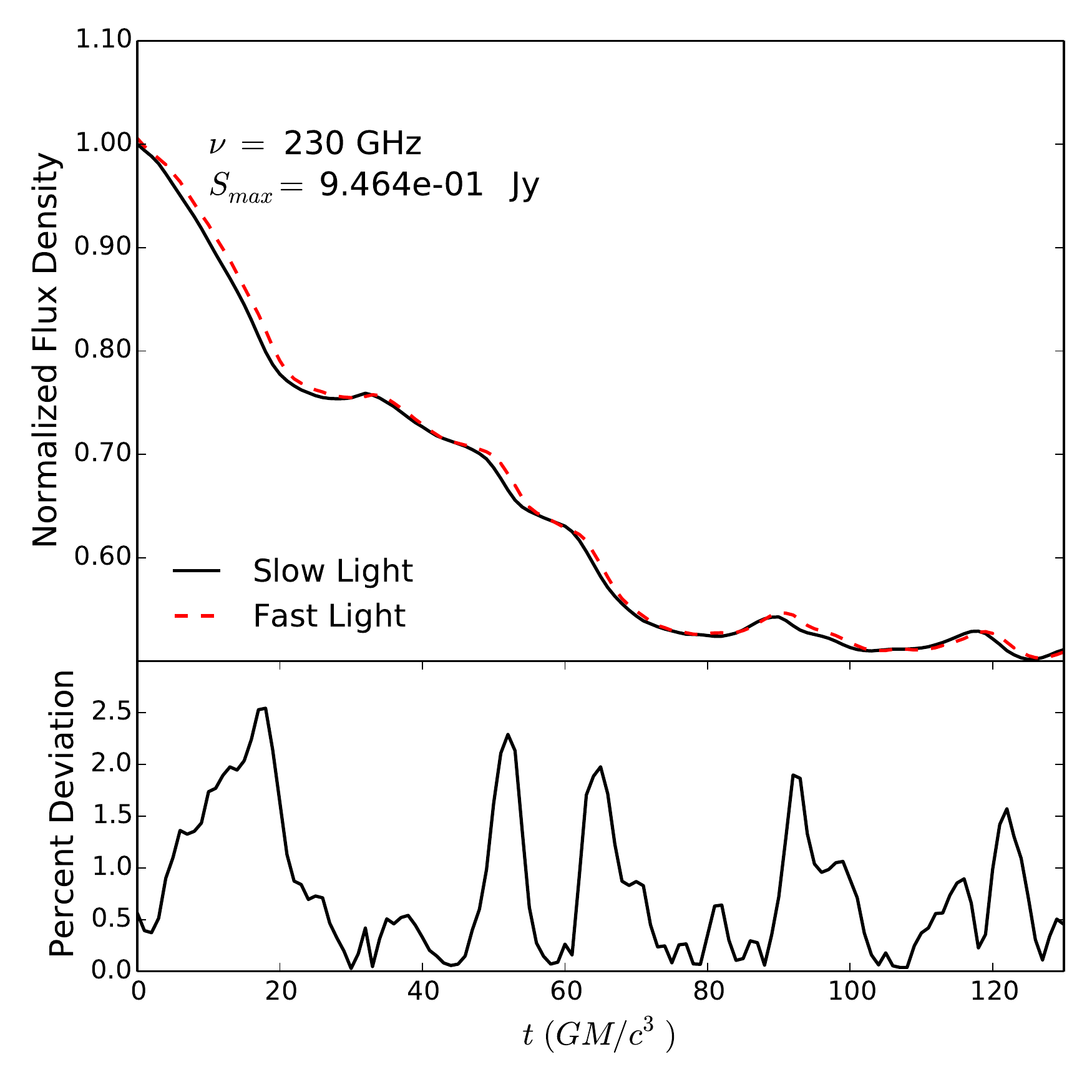} 
\includegraphics[width=0.45\textwidth]{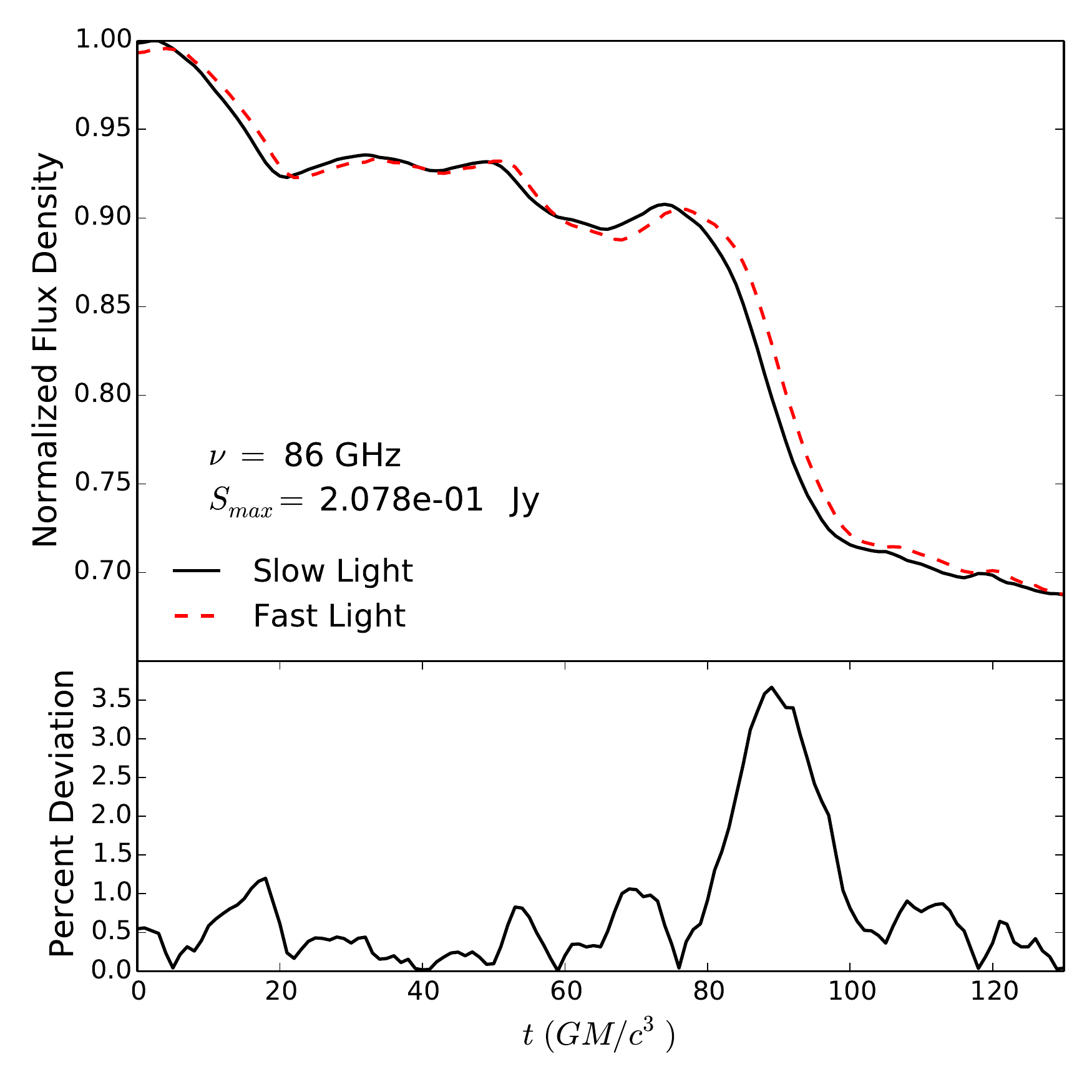}\\
\includegraphics[width=0.45\textwidth]{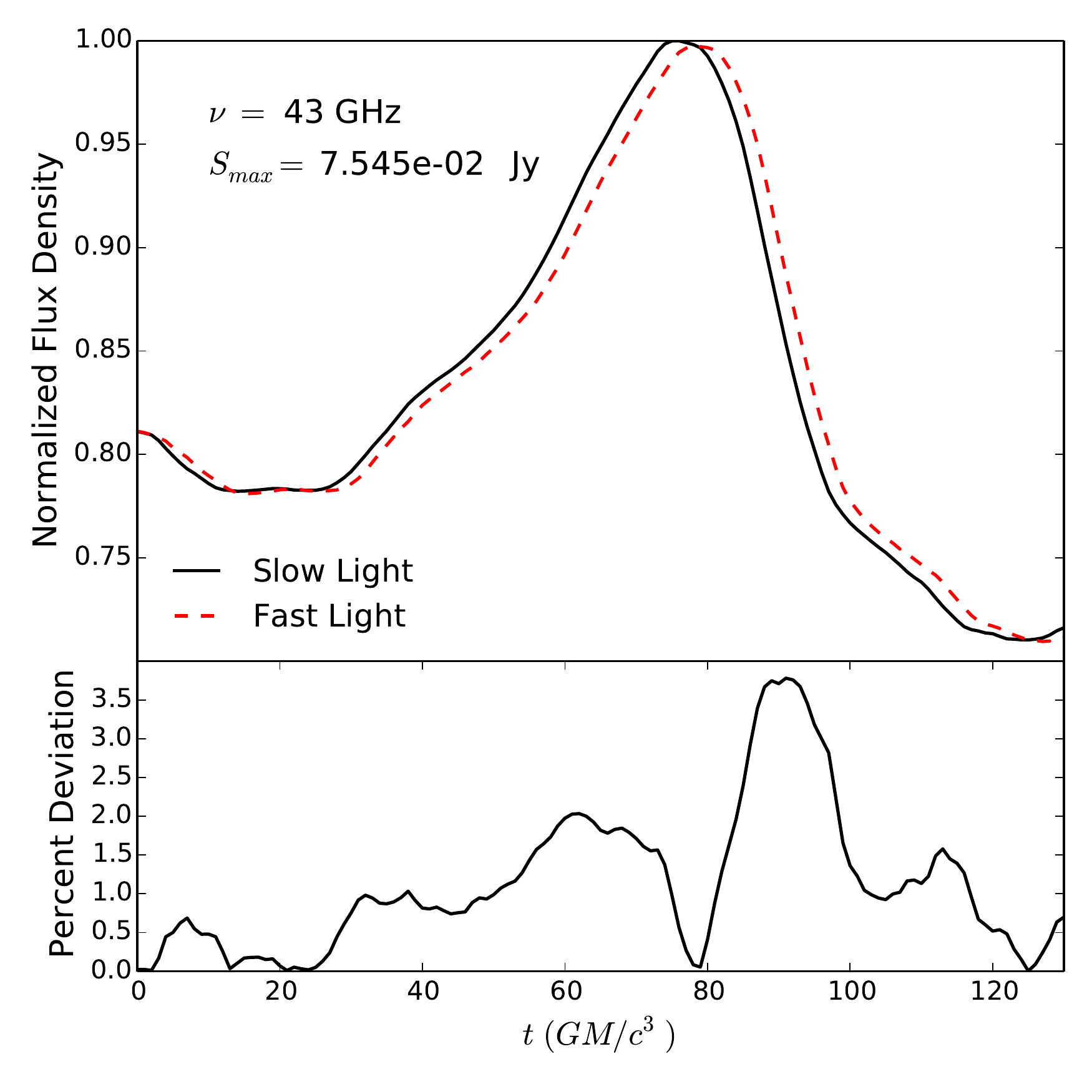}
\includegraphics[width=0.45\textwidth]{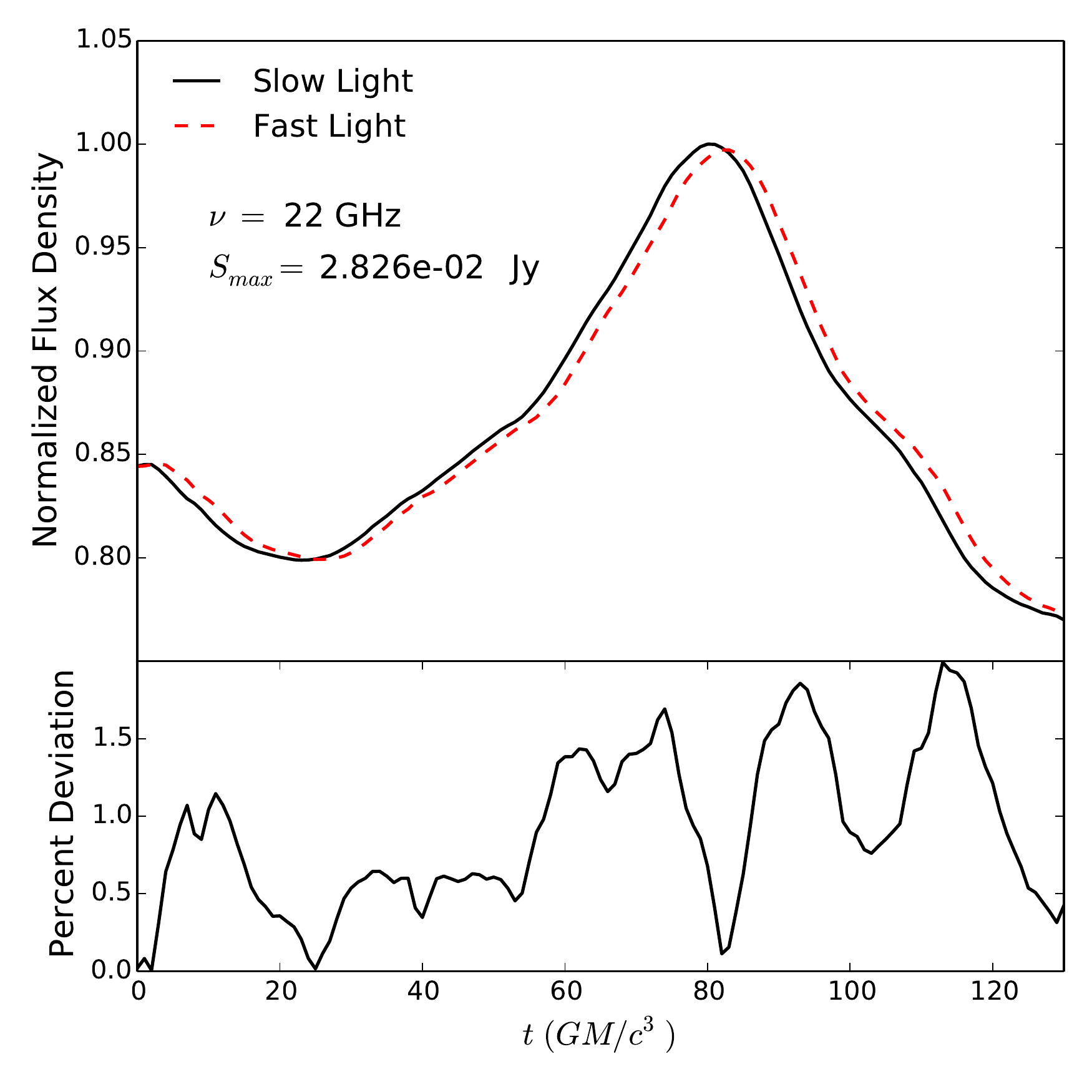}
\end{center}
\caption{Light curves for slow-light (solid) and fast-light (dashed) for
  the jet emission dominated model at VLBI frequencies. Residuals are
  displayed below the light curves.}
	\label{fig:lightcurves_jet_230GHz_dashed_FL2}
\end{figure*}

\begin{figure*}
\includegraphics[width=\textwidth]{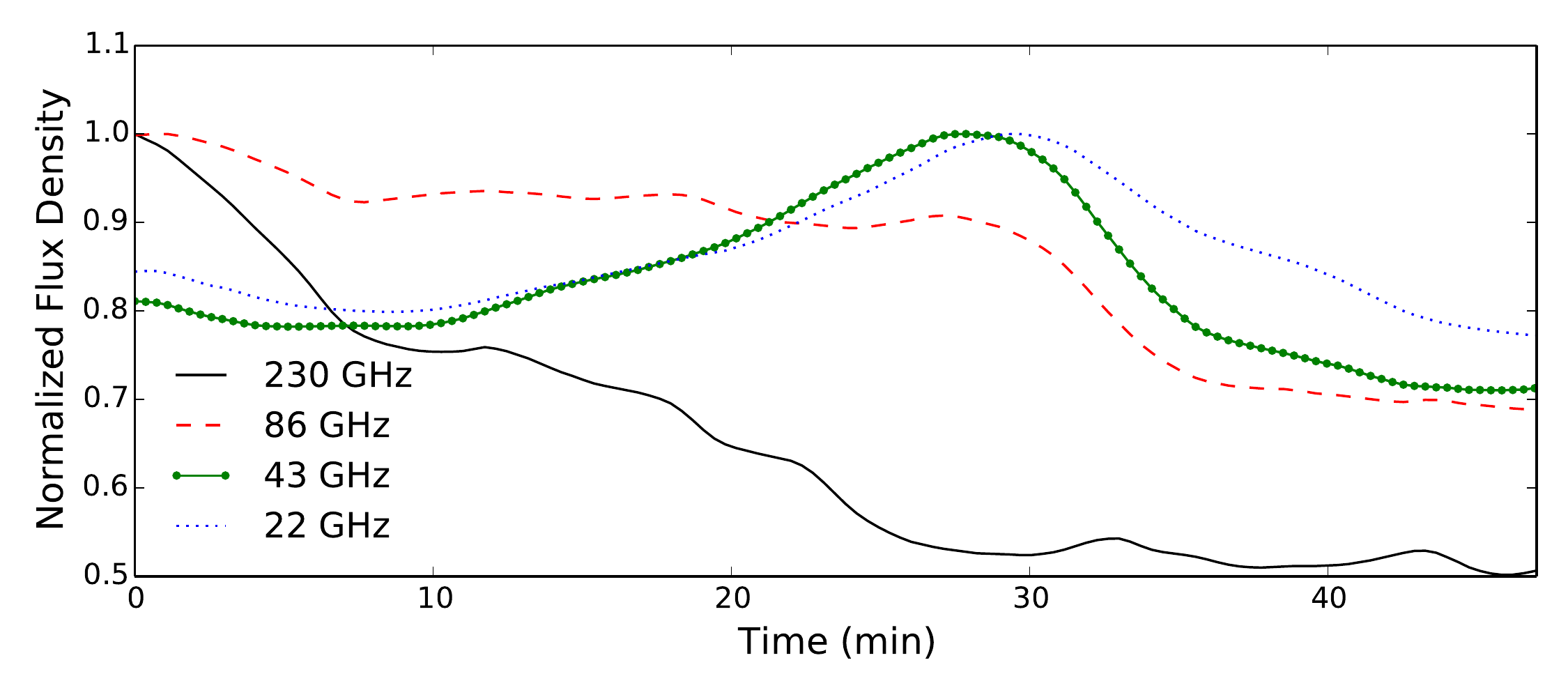}
\caption{Normalized light curves of slow-light simulations at the VLBI
  frequencies (230 GHz (solid), 86 GHz (dashed), 43 GHz (solid and
  dotted), 22 GHz (dotted)) for viewing angle of $60\degr$ for our jet
  dominated model.}~\label{fig:multifreq60deg_jet}
\end{figure*}

Similarly, the model whose emission is dominated by the jet is
characterized by a mass scale factor $\mathcal{M}=2.515 \times 10^{-13} M_\odot$ and the weak-magnetization
electron-proton temperature ratio $R_{\rm high}=25$ (while $R_{\rm low}=1$).  In this case,
slow-light images of the accretion flow at all VLBI frequencies are shown
in Fig.~\ref{fig:230GHz_jet_SL}, while the light curves of fast- and
slow-light simulations are reported in
Fig.~\ref{fig:lightcurves_jet_230GHz_dashed_FL2}; finally the normalized
slow-light light curves at all frequencies are shown in
Fig.~\ref{fig:multifreq60deg_jet}.

We found that the difference between the fast-light and slow-light approaches
is systematically below 5\% in all cases, suggesting that the fast-light
approximation is a good one for the model presently under
consideration, a result that is in line with the findings of
\citet{dexter2010}.

As a concluding remark, we note that we have considered radiative-transfer
calculations of the unpolarized component of synchrotron radiation. This
is motivated by the fact that observations show that linear and circular
polarization of Sgr~A* are less than a few percent
\citep{bower2003}. Although small, the polarization is nonzero and we
here speculate that the differences between the two approaches may become
more pronounced when considering the polarized components of the
radiation.

\section{Summary}

We have introduced {\tt RAPTOR}, a new time-dependent radiative-transfer
code capable of producing physically accurate images of black-hole
accretion disks as calculated by GRMHD simulations in particular, and of performing radiative-transfer calculations in general astrophysical problems that involve radiative transfer and strong gravity in general. The code achieves
this result by integrating null geodesics in arbitrary spacetimes and
then performing time-dependent radiative transport along the geodesics to
ultimately construct images, light curves, and spectra. Furthermore, {\tt
  RAPTOR} is capable of time-dependent computations, in which the
characteristics of a single image may depend on a range of input data in
the time domain and it can be run on both CPUs and GPUs.

We have verified correctness of our geodesic calculations by
investigating the conservation of constants of motion and comparing our
results to the semi-analytical code {\tt geokerr}, our results being in
good agreement. The same procedure was applied to our radiative-transfer
computations: we performed a number of tests of our various integration
schemes and reproduced results from the literature.

As an additional verification step, we compared our code directly to the
another recently developed radiative-transfer code: {\tt BHOSS}
\citep{BHOSS}. We considered a complex scenario that involves a
particular set of GRMHD data along with a physical radiative model and
produced images of the accretion flows using both codes. The result was a convergence of the total
flux density as computed by the two codes, which rely on different integration
algorithms for both null geodesics and radiative transfer, to a relative error of less
than 0.01\% (See Fig.~\ref{fig:BHOSSRAPTOR}), demonstrating that our implementations are highly consistent.

An important validation of our code has come from neglecting a commonly
made assumption that the physical properties of the underlying plasma do
not vary during the propagation of the radiation (fast-light
approximation). We have therefore studied the difference between the
slow-light and fast-light paradigms in radiative models of accretion
flows onto SMBHs. The differences between slow-light and fast-light can
be arbitrarily large or utterly negligible, depending on the physical
conditions. In context we have considered, where we have investigated
radiative models based on 2D GRMHD data, we conclude that the effects of
switching between the fast-light and slow-light paradigms on the
appearance of this particular model's accretion flow are smaller than 5\%
across the VLBI frequencies we examined. 

While these results are in agreement with the findings of
\citet{dexter2010} we note that this conclusion may be
 influenced by the fact that we have investigated unpolarized radiative transfer only.
In conclusion, the {\tt RAPTOR} code -- and its further developments in terms of the
possibility of treating polarized radiative transfer -- may be used to
compare synthetic images of SMBH accretion flows to mm-VLBI data from the
Event Horizon Telescope Collaboration
(\url{http://eventhorizontelescope.org}) and to extract physical
properties of the black-hole system (e.g., spin, inclination) or to test
the predictions of GR about the the size and shape of the black-hole
shadow. The code has many potential applications in other astrophysical problems, such as radiative transfer near a neutron star or a binary black-hole system. It is also well-suited to creating material for outreach purposes, such as virtual-reality movies of black hole environments \citep{vrpaper}.

Finally, we note that because the electron distribution function of the
plasma is an important factor in determining radiative properties,
reducing the number of assumptions made about this function may have an
appreciable effect on the synthetic observational data one may generate
using our models. In view of this, we have recently added a more general
particle distribution function than the one presented here, relaxing the
assumption that the electrons are in a thermal distribution by adding
accelerated particles and thereby improving the accuracy of radiative
models. The results of this analysis will be presented in a distinct and
forthcoming paper \citep{accpart}.


\section*{Acknowledgements}

This work is supported by the ERC Synergy Grant "BlackHoleCam: Imaging
the Event Horizon of Black Holes" (Grant 610058). Z.Y. acknowledges
support from an Alexander von Humboldt Fellowship. This research has made
use of NASA's Astrophysics Data System. Some of the simulations in this work were performed on
the LOEWE cluster in CSC in Frankfurt.

\bibliography{BHC}
\bibliographystyle{apalike}

%


\begin{appendix} 

\section{Coordinate Transformations}
\label{appB}

Here, we list certain non-trivial coordinate transformations that are
used in {\tt RAPTOR}. In what follows, the BL coordinates are denoted by
$(t,r,\theta,\phi)$, and the KS coordinates by
$(\tilde{t},r,\theta,\tilde{\phi})$.

\subsection{BL-KS}

The transformations of the coordinate vector between BL and KS
coordinates is given by (see e.g., \citet{font1999}):
\begin{eqnarray}
\tilde{t} &=& t + M\ln\Delta + \frac{2M^{2}}{r_{+}-r_{-}}\ln \left(
\frac{r-r_{+}}{r-r_{-}} \right) \ , \\ \tilde{\phi} &=& \phi +
\frac{a}{r_{+}-r_{-}}\ln \left( \frac{r-r_{+}}{r-r_{-}} \right) \ ,
\\ \dot{\tilde{t}} &=& \dot{t} + \frac{2Mr}{\Delta}\dot{r} \ ,
\\ \dot{\tilde{\phi}} &=& \dot{\phi} + \frac{a}{\Delta} \dot{r} \ ,
\end{eqnarray}
where an overdot denotes differentiation with respect to the affine
parameter, $\lambda$, $r_{\pm}:= M \pm \sqrt{M^{2}-a^{2}}$ denotes
the outer(inner) event horizon radius, $\Delta = r^{2}-2Mr+a^{2}:=
(r-r_{+})(r-r_{-})$ and $M$ is the black-hole mass. It is important to note that these
transformations are valid only in the region of spacetime exterior to the
black-hole horizon.

Four-vectors transform differently. In {\tt RAPTOR}, the initial
contravariant wave vector $k^{\mu}_0$ is always constructed using BL
coordinates, and must be transformed to KS coordinates. This is
accompanied by the following transformation matrix \citep{mckinney2004}:
\begin{equation}
k^{\bar{\alpha}} = 
\renewcommand{\arraystretch}{1.2}
    \left(
         \begin{array}{cccc}
1 & 2r/\Delta & 0 & 0 \\
0 & 1 & 0 &  0 \\
0 & 0 &1 & 0 \\
0& a/\Delta& 0 & 1    
        \end{array}
    \right) k^{{\alpha}}\,,
\end{equation}

where $k^{\alpha}$ denotes the wave vector in BL coordinates,
$k^{\bar{\alpha}}$ is the wave vector in KS coordinates. The reverse
transformation is given by
\begin{equation}
k^{{\alpha}} = 
\renewcommand{\arraystretch}{1.2}
    \left(
         \begin{array}{cccc}
1 & -2r/\Delta & 0 & 0 \\
0 & 1 & 0 &  0 \\
0 & 0 &1 & 0 \\
0& -a/\Delta& 0 & 1    
        \end{array}
    \right) k^{\bar{\alpha}}\,.
\end{equation}

\subsection{KS-MKS}

The modified Kerr-Schild coordinates (MKS), as used by, for example, \cite{gammie2003}, are denoted by
$(\tilde{t},x^{1},x^{2},\tilde{\phi})$. Some of these quantities may be
expressed in terms of conventional KS coordinates as:
\begin{eqnarray}
x^{1} &=& \ln \left( r - r_{0} \right) \ , \label{x1} \\ \dot{x}^{1} &=&
\frac{\dot{r}}{r-r_{0}} \ , \label{xdot1} \\ \dot{x}^{2} &=&
\frac{\dot{\theta}}{ \pi \left[ 1 + \left( 1 - h \right) \cos
    \left( 2\pi x^{2} \right) \right]} \ , \label{xdot2}
\end{eqnarray}
where $0\le h \le 1$ is obtained from the GRMHD data and stretches the
zenith coordinate near the poles and the equatorial plane. Notice that
we cannot obtain $x^{2}(\theta)$ algebraically as this requires solving a
transcendental equation; we must find it numerically, via the inverse
transformation $x^2 \rightarrow \theta$, which can be written in
closed form. This transformation, along with the corresponding inverse
transformations for Eqs.~(\ref{x1}), (\ref{xdot1}), and (\ref{xdot2}),
may be written as:
\begin{eqnarray}
r &=& r_{0} + \exp{x^{1}} \ , \label{r_MKS} \\ \theta &=& \pi x^{2} +
\frac{1}{2}\left( 1-h \right)\sin \left( 2\pi x^{2} \right)
\ , \label{theta_MKS} \\ \dot{r} &=& \dot{x}^{1} \left( r - r_{0} \right)
\ , \label{rdot_MKS} \\ \dot{\theta} &=& \pi \dot{x}^{2}\left[ 1 +
  \left(1-h \right)\cos \left(2\pi x^{2} \right) \right]
\ . \label{thetadot_MKS}
\end{eqnarray}

To perform the transformation $\theta \rightarrow x^2$, we must resort to
numerically (indeed iteratively) seeking a value $x^2$ that satisfies
Eq.~\eqref{theta_MKS}, and this is readily performed using the
Newton-Raphson method. For a function $f(x)$ the solution to $f(x)=0$
may be found iteratively as:
\begin{equation}
x_{n+1} = x_{n} - \frac{f(x_{n})}{f'(x_{n})} \,,
\end{equation}
where $f'(x_{n}):=\mathrm{d}f(x_{n})/\mathrm{d}x_{n}$ and the index
$n$ denotes the iterative step. For MKS coordinates we have:
\begin{eqnarray}
f(x^{2}_{n}) &=& \left[\pi x^{2}_{n} + \frac{1}{2}\left( 1-h
  \right)\sin \left( 2\pi x^{2}_{n} \right) - \theta \right]
\ , \\ f'(x^{2}_{n}) &=& \pi \left[ 1 + \left(1-h \right)\cos
  \left(2\pi x^{2}_{n} \right) \right] \,,
\end{eqnarray}

since $\partial\theta/\partial x^{2}_{n}=0$ and $\theta$ is constant in
the above iterative scheme. One must start with a trial value for
$x^{2}_{0}\ll 1$ and then iterate from there. Since $x^{2}\in [0,1]$, if
$x^{2}_{n+1}<0$ or $x^{2}_{n+1}>1$ then reset $x^{2}_{n+1}$ to a small
value (e.g., $0.05$). We also apply the above reset if $f'(x^{2}_{n})
\rightarrow 0$. Finally, we define a appropriate convergence criterion,
namely $|x^{2}_{n+1} - x^{2}_{n}| < \varepsilon$, where $\varepsilon$ is
the error tolerance we find acceptable.

\section{Code performance}
\label{sec:code_performance}

As mentioned in the Introduction, an important added value of {\tt
  RAPTOR} is that it is a hybrid code that can be compiled for both CPUs
and GPUs. Two distinct parallelization methods are implemented in the
code; one is {\tt OpenMP}\footnote{\url{http://www.openmp.org}}, with
which it is possible to run the code on multiple CPU cores; the other is
{\tt OpenACC}, with which it is possible to run the code on both the CPU
and GPU. We have verified and tested the performance of the code in these
environments by considering a radiative-transfer calculation through our
{\tt BHAC} GRMHD simulation. The model parameters are listed in
Table~\ref{table:perf-setup}. More specifically, we investigate how the
performance of {\tt RAPTOR} scales with the number of threads that are
employed. This is done by running the same setup
(Table~\ref{table:perf-setup}) on the same hardware
(Table~\ref{table:hardware}) multiple times, using a different number of
cores each time.

The scalability is measured by calculating the speed-up factor, which is
defined as
\begin{equation}
{\rm S} = T_{\rm \#threads} / T_{\rm single\ thread}\,,
\end{equation}
where $T$ is the run time of the code for a given amount of threads.

\begin{figure}
\includegraphics[width=0.24\textwidth]{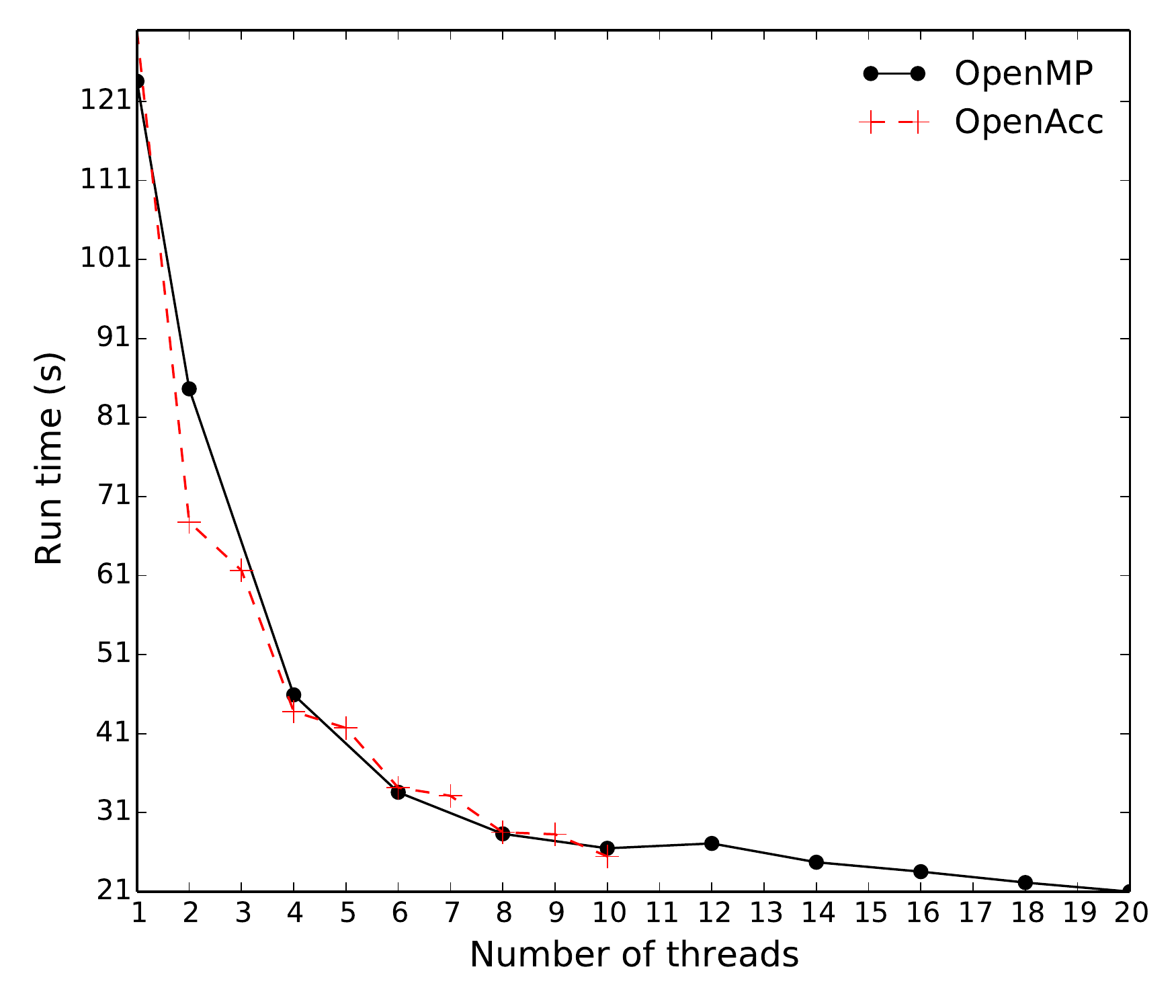}
\includegraphics[width=0.24\textwidth]{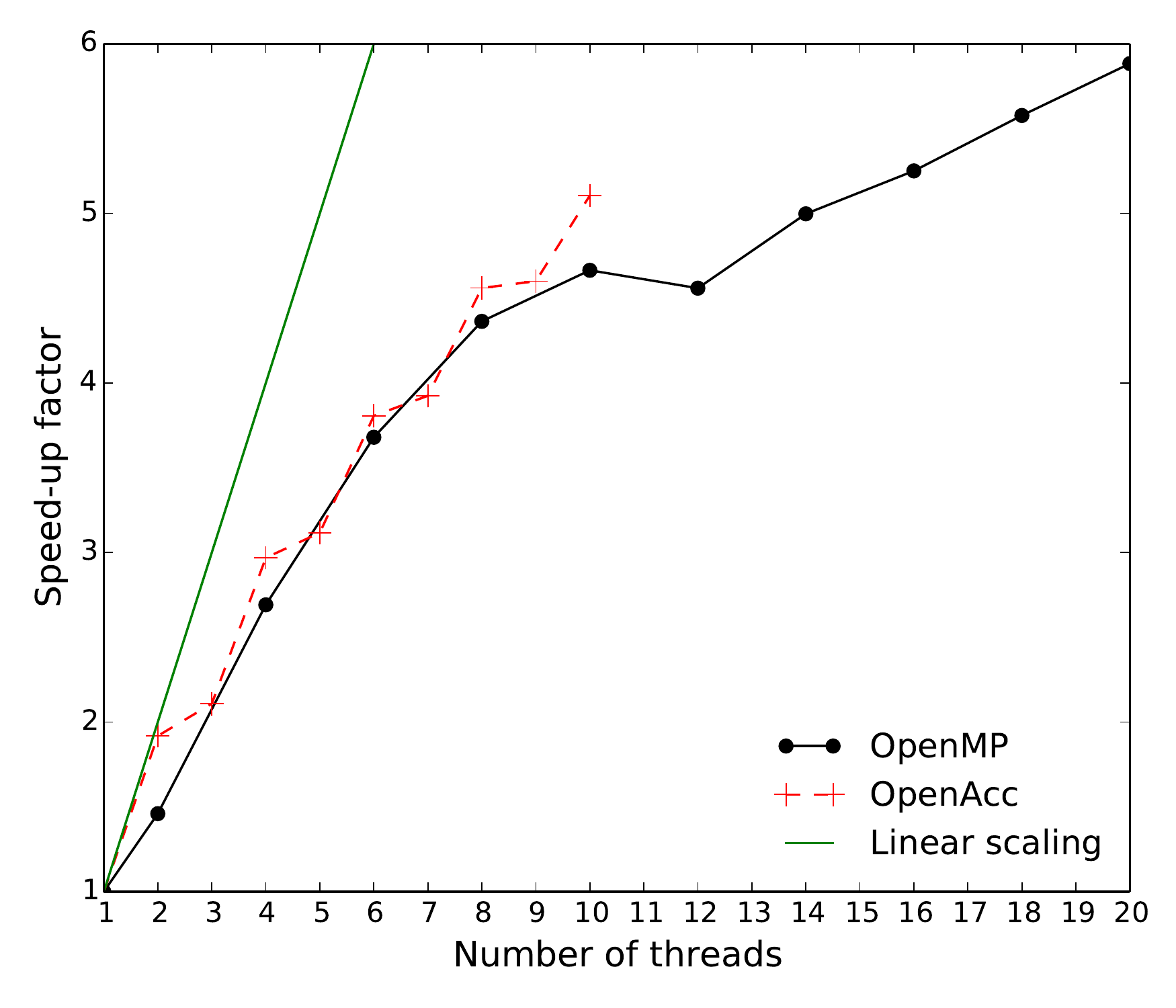}
\caption{Run time (left) and speed-up factor (right) for {\tt RAPTOR}
  using {\tt OpenMP} and {\tt OpenACC}.}
\label{fig:runtime}
\end{figure}

The results of our runs can be found in Fig. \ref{fig:runtime} and they
show that the code scales sub-linearly with the number of cores, as is
expected, since communication overhead increases with an increasing
number of threads; again, as expected, the GPU runs outperform the CPU
ones. Interestingly, the {\tt OpenMP} implementation slightly outperforms
{\tt OpenACC}, but this is not surprising, since {\tt OpenACC} lacks
hyper-threading support for CPUs, which is instead provided with {\tt
  OpenMP}. When using 10 CPU-cores, {\tt RAPTOR} can integrate 10,707
geodesics per second, while with one GPU unit, {\tt RAPTOR} integrates
104,900 geodesics per second.


\begin{table}
\centering
\begin{tabular}{l l}
\hline
\hline
 Parameter & Value  \\
\hline
Single precision & yes\\
Amount of pixels & $512 \times 512$\\
Image size & $40\,M$\\
step-size $\epsilon$ (Eq.~\eqref{adaptivestep_1}) &   0.03 \\
Integration scheme  & RK2 \\
Radiative transport & yes\\
\hline
\end{tabular}
\caption{Numerical parameters for the code performance tests.}
\label{table:perf-setup}
\end{table}

One reason why the difference between GPU and CPU performance is
relatively small is the additional time required for data transfer and
kernel booting operations in the GPU based implementation. We therefore
also rendered a large image ($2000\times 2000$ pixels) on both
the CPU and GPU, to check whether the difference between CPU and GPU
performance increases. The average run times are of 3 min, 58 sec on one
CPU and of and 39 sec one GPU, respectively. As expected, the performance
difference is substantially increased under these conditions. These total
run times also include data read ($\sim1$ second) and output generation
($\sim0.2$ seconds) operations.

\begin{table}
\centering
\begin{tabular}{l l l}
\hline
\hline
 & {\tt OpenMP} & {\tt OpenACC}  \\
\hline
CPU & Intel i7-6950X  & Intel i7-6950X CPU\\
No. cores & 10 & 10\\
Multi-threading & yes & no \\
Clock speed & 3.0 GHz & 3.0 GHz\\
GPU & - & GeForce GTX 1080 \\
No. CUDA cores & - & 2560 \\
Compiler & gcc & pgcc \\
Optimization flags & -O3 & fastmath \\
\hline
\end{tabular}
\caption{Description of the hardware on which our performance tests were
  executed.}
\label{table:hardware}
\end{table}

\end{appendix}

\end{document}